\documentclass[1p]{elsarticle}

\usepackage{setspace}
\usepackage{xcolor}
\usepackage{graphicx}
\usepackage[labelfont=bf,labelsep=period,skip = 0pt]{caption}
\usepackage{xr-hyper}
\usepackage{chemformula} 
\usepackage[T1]{fontenc} 
\usepackage{multirow}
\usepackage{comment}
\usepackage{siunitx}
\usepackage{amssymb}
\usepackage[normalem]{ulem}
\usepackage{xr}
\makeatletter
\newcommand*{\addFileDependency}[1]{
	\typeout{(#1)}
	\@addtofilelist{#1}
	\IfFileExists{#1}{}{\typeout{No file #1.}}
}
\makeatother

\newcommand*{\myexternaldocument}[1]{
	\externaldocument{#1}
	\addFileDependency{#1.tex}
	\addFileDependency{#1.aux}
}
\myexternaldocument{supporting}
\biboptions{numbers,sort&compress}

\journal{Chemical Engineering Journal}

\begin{document}
\begin{doublespace}
\begin{frontmatter}

\title{Asphaltene precipitation under controlled mixing conditions in a microchamber}

\author[First]{Jia Meng}
\author[First,Second]{Chiranjeevi Kanike}
\author[First]{Somasekhara Goud Sontti}
\author[Second]{Arnab Atta}
\author[First]{Xiaoli Tan\corref{cor1}}
\ead{xiaolit@ualberta.ca}
\author[First]{Xuehua Zhang\corref{cor2}}
\ead{xuehua.zhang@ualberta.ca}
\address[First]{Department of Chemical and Materials Engineering, University of Alberta, Alberta T6G 1H9, Canada}
\address[Second]{Department of Chemical Engineering, Indian Institute of Technology Kharagpur, Kharagpur, West Bengal 721302, India}
\cortext[cor1]{Corresponding author}
\cortext[cor2]{Corresponding author}

\begin{abstract}
Solvent exchange is a controlled process for dilution-induced phase separation.
This work utilizes the solvent exchange method to reveal the effect of the mixing dynamics on the asphaltene precipitation process under 20 different mixing conditions using a model system of n-heptane and asphaltene in toluene. The external mixing between the asphaltene solution and the paraffinic solvent is strictly controlled. We employed high-spatial resolution total internal reflection fluorescence microscope to detect asphaltene precipitates with a resolution up to $\sim$ 200 $nm$. A multiphysics model is used to simulate the evolution of oversaturation pulse in the solvent exchange process. Based on the simulation results, we predicted the effect of the flow rate, dimension, orientation of the microfluidic chamber, and temperature on the surface coverage and size distribution of asphaltene precipitates. The model predictions of all factors corroborate with the experimental observations. Local concentration of the solvent and shear forces are found to be the two main reasons for the change of asphaltene precipitation caused by mixing dynamics. However, the influence of thermodynamics is more critical than the mixing dynamics as temperature changes. Through a combination of experimental and simulation studies, this work illuminates the significance of the transportation process for the final morphology of asphaltene precipitates and provides a in-depth insight into the mechanism of mixing dynamics on the asphaltene precipitation. A smart mixing may be to boost new phase formation without excessive solvent consumption.

\end{abstract}


\end{frontmatter}


\section{Introduction}
Asphaltene plays an important role in energy sector through production, refinery, and transportant to value-added product from hydrocarbon-rich waste \cite{gray2015upgrading}. As a complex mixture containing polycyclic aromatic hydrocarbons, asphaltene is defined by solubility class that is soluble in aromatic solvents such as benzene and toluene and insoluble in n-alkanes solvents such as n-pentane and n-heptane \cite{schuler2015unraveling}. On one hand, the presence of asphaltene hinders the coalescence of oil-water emulsions \cite{mao2019novel,ma2020novel}, causing increase in operational cost and production challenges by forming flocculation and deposits in the reservoirs and transportation pipes. Undesirable asphaltene precipitation in the transportation pipes can cause severe problems such as clogging and blocking \cite{keshmiri2019microfluidic,zi2021investigation}. \textcolor{black}{On the other hand, dilution-induced asphaltene precipitation has been utilized in paraffinic froth treatment (PFT) units to remove impurities in bitumen such as water and solids. In this process, the particle size of asphaltene precipitates plays an important role, as they interact and agglomerate with water droplets and solid particles \cite{rao2013froth}.} The precipitated asphaltene can be used as a raw material to make value-added products, such as absorbent, or carbon fibers \cite{plata2022characterization,tehrani2019novel,alivand2019tuning,saad2022transformation}. Therefore, it is of importance to understand and control asphaltene precipitation. 

Asphaltene is considered to form nano-aggregates and nano-clusters in a crude oil or aromatic solvents, unless at high dilution \cite{yarranton2013size,mullins2010modified,elkhatib2019nanoscale,zhao2007composition,eyssautier2012mesoscale}. Asphaltene dispersed in toluene is more stable due to the balance of attractions (van der Waals, H-bonding, and acid base interactions) and steric repulsions originated from side alkane chains \cite{gray2011supramolecular,da2012density}. Adding n-alkanes such as paraffinic solvents into the mixture of asphaltene and toluene leads to the collapse of the side alkane chain layer and a reduction in steric repulsion. Consequently, asphaltene nano-aggregates grow to a larger size and manifest as precipitates or aggregates \cite{wang2010interaction,wang2009colloidal}. The effective parameters that may affect asphaltene precipitation include temperature \cite{peramanu1999flow}, pressure \cite{akbarzadeh2005generalized}, addition of inhibitors \cite{alemi2021experimental}, solvent to bitumen (S/B) ratio \cite{rogel2017effect,hristova2022asphaltene}, and type of the solvent \cite{alboudwarej2003regular}. 

Recently, the effect of mixing conditions on asphaltene precipitation has attracted the attention of researchers to understand aggregate formation \cite{rahmani2003characterization, rahmani2004evolution,boek2010multi,boek2008deposition,ali1998effect}. In a conventional bulk system, when crude oil and paraffinic solvents were added into a container and mixed by a mixer, the resulting shear force caused by the agitation enhanced the growth rate of asphaltene particles, but the final size of the asphaltene particles was found to be smaller \cite{rahmani2003characterization, rahmani2004evolution}. The increase of the flow rate helped to increase the rate of deposition but the thickness of the final deposit layer was reduced \cite{boek2010multi,boek2008deposition,ali1998effect}. 

The influence of mixing dynamics on asphaltene precipitation is mainly from two aspects. The impact of shear force on asphaltene aggregation and fragmentation \cite{rahmani2005online,rastegari2004kinetics,elkhatib2019nanoscale,nguyen2021effect} and the variation in local compositions triggering asphaltene precipitation \cite{shalygin2019spectroscopic,morozov2017reversibility,buckley2006mixing,durand2010effect}. Extensive studies have been carried out to predict the deposition. However, there are knowledge gaps in the dynamics of dilution-induced asphaltene precipitation at the early stage. Importantly, the effects of local flow concentration and convection on asphaltene precipitation are challenging due to the poor mixing control. On the other side, the sampling procedures make it impossible to control convection. The results obtained from the direct visualization of asphaltene particles at a submicron scale may not be sufficient to investigate early-stage asphaltene precipitation due to the limited resolution of conventional optical microscopes. Microfluidic devices have been applied in the study of asphaltene precipitation with high reproducibility of results \cite{mozaffari2021lab,pagan2022physicochemical}. For example, understanding the characteristics of the precipitates from mixing in confined spaces \cite{meng2021primary,meng2021microfluidic,meng2022size}. 

Solvent exchange, also called solvent shifting or solvent displacement has shown promising results in a quantitative understanding of solvent-induced phase separation \cite{Zhang9253}. Controlled mixing by solvent exchange has led to fundamental insights into phase separation induced by dilution, including nanobubble and nanodroplet formation \cite{peng2018growth,meng2020viscosity}, oiling out crystallization \cite{zhang2021oiling,choi2021effects}, and polymer nanoprecipitation \cite{zhang2012flash}. The approach of solvent exchange is displacing a solution in a good solvent with a poor solvent. Consequently, the new phase nucleates and grows at the mixing front due to a transient oversaturation \cite{dyett2018growth}. The solvent exchange process is a well-controlled mixing system, which can be described by the advective diffusion equation \cite{Zhang9253}. Overall, the diffusive and convective mixing in the system is controllable by varying the dimension of the microfluidic chamber and the injection rate of the solvents. To the best of our knowledge, such fundamental studies on asphaltene precipitation are not yet reported in the open literature. In particular, the orientation and channel height of the chamber, the flow rate of the mixing, and the operating temperature can dictate the final morphology of asphaltene precipitates in terms of surface coverage ($SC$) and size distribution at the submicron and micron scales. 

This research aims to understand the solvent-induced asphaltene precipitation under controlled convective mixing. The microfluidic flow rate, microchannel height, and orientation all enable the control of local concentration of the solvent and local shear force in the experiments. Systematic numerical investigations based on the finite element method (FEM) were performed to visualize the oversaturation pulse in the solvent exchange process for different geometric and flow conditions. Through the numerical results, we identified the generation of convective rolls when the channel height was more than a critical value due to gravity and buoyancy forces. Furthermore, the total internal reflection fluorescence microscope (TIRF) and optical microscope are used to resolve the asphaltene particles at the submicron and micron scales. In this work, the surface coverage ($SC$) is characterized to quantify the asphaltene yield. The size distribution at the submicron and micron scales is specified to investigate the aggregation of asphaltene particles. The implication is that the optimized flow condition may mediate the asphaltene precipitation without leveraging solvents. The findings of this work \textcolor{black}{provide} new insights into the influence of mixing condition on asphaltene precipitation, and is not only applicable to asphaltene precipitation but also has implications for other dilution-induced phase separation processes.

\section{Experimental section}
\subsection{Chemical and substrate preparation}
Murphy Oil (USA) provided the C5-asphaltene (i.e., n-pentane precipitated asphaltene) used in this project. The analysis of elementary and the pre-treated method of the asphaltene sample can be found in our previous research \cite{MENG2021120584}. In brief, raw asphaltene sample was dissolved in toluene (Fisher Scientific, ACS grade, 99.9 \%+) to remove any inorganic solvents by 0.22 $\mu m$ polyvinylidene difluoride (PVDF) filter paper. A rotary evaporator was then used to remove toluene from the inorganic solid free asphaltene. 17 $g$ of inorganic solid-free asphaltene sample was dissolved in toluene to make 1 $L$ of asphaltene in toluene solution (i.e., solution A). \textcolor{black}{The results of diluted crude oil are similar to that of asphaltene in toluene solution at early stage of asphaltene precipitation \cite{meng2022size}.} n-Heptane (Fisher Chemical, 99\%) was used as the solvent (i.e., solution B) for asphaltene precipitation.

A high-precision cover glass (Azer Scientific) was used as the substrate. Before the experiment, the cover glass was sonicated in toluene for 5 min and dried by air to remove any contaminates.

\subsection{Solvent exchange process}
\begin{figure}[ht]
\centering
\includegraphics[width=1\columnwidth]{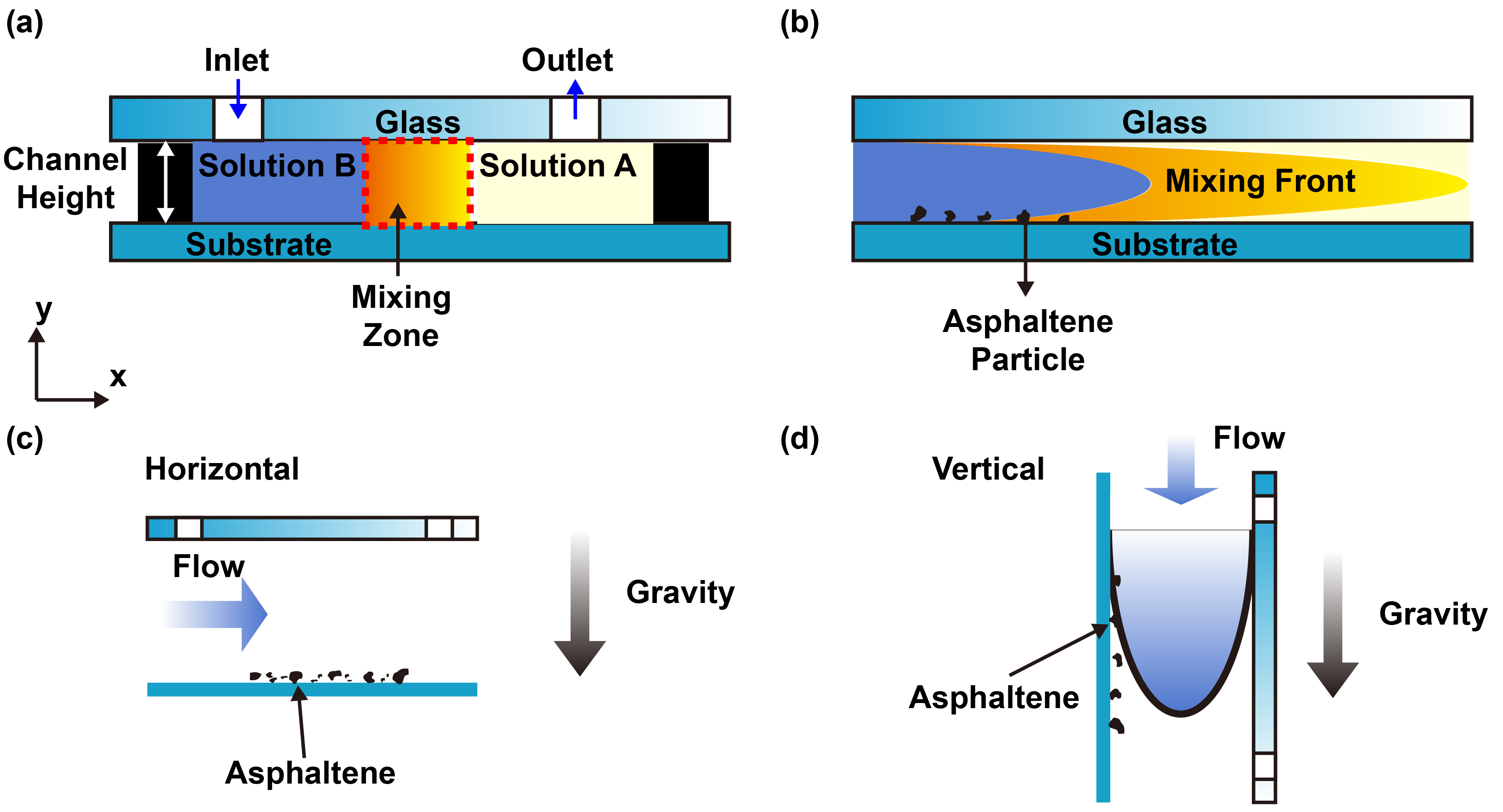}
\renewcommand{\captionfont}{\linespread{1.6}\normalsize}
\caption{(a) Sketch of the microchamber used in this study. (b) Zoomed-in image of the mixing front in (a). The flow direction is along the x-axis. The microchamber is placed (c) horizontally, (d) vertically to examine the effect of orientation.}
\label{chamber_5}
\end{figure}

The solvent exchange process was performed in a custom-built microchamber. \textcolor{black}{The experiments were performed at ambient pressure, as the current design of microfluidic chamber did not allow for varying the pressure.} The apparatus is shown in Figure \ref{chamber_5}(a). The micro-channel was 35 $mm$ long and 5 $mm$ wide. Glass was used as a top wall and a spacer in between the glass and substrate was used to seal the system. The channel height was adjusted by using spacers of different thickness.

The model oil (solution A) was first injected into the microchamber using a syringe pump (NE-1000, Pumpsystems Inc.). Afterwards, n-heptane (solution B) was injected into the device to displace solution A, creating a mixing front between solution A and solution B as shown in Figure \ref{chamber_5}(b). The asphaltene precipitation began at the moment when the concentration of n-heptane is higher than the onset point in the mixing front. At the moment, the solution A was completely displaced by solution B, asphaltene precipitation stopped due to lack of dissolved asphaltene. After injecting 2.5 $mL$ of solution B (i.e., at least ten times the total volume of solution A), the asphaltene precipitation was considered the final state. Air was injected into the chamber to drive away from the remaining liquid. After the solvent exchange process, the liquid was removed with a 5 $mm/s$ of a constant air flow velocity, in all of the experimental conditions. Subsequently, the substrate was examined by the microscopes. The microchamber was placed both horizontally (Figure~\ref{chamber_5}(c)) and vertically (Figure~\ref{chamber_5}(d)) to examine the gravitational effect.


\subsection{Effect of flow rate, channel height, orientation, and temperature}
This work investigated the influence of four aspects on asphaltene precipitation that include injection flow rate of solution B, channel height, orientation of the microfluidic chamber, and temperature. A horizontally placed 1,000 $\mu m$ height microfluidic chamber was used to study the influence of flow rate under the ambient condition (i.e., 19 - 21$^{\circ}$C, 1 $atm$). The flow rate of solution B was varied from 15 $mL/h$ to 150 $mL/h$. The Reynolds number ($Re$) and P\'{e}clet number ($Pe$) were calculated from $Re = \frac{Q}{w \upsilon}$ and $Pe = \frac{Q}{wD}$, respectively \cite{Zhang9253}, where $Q$, $w$, $\upsilon$, $D$ are the microfluidic flow rate, width of the microchamber, kinematic viscosity, and diffusion coefficient, respectively. Table \ref{table 2} summarizes the flow rate and the corresponding $Re$ and $Pe$. 

\begin{table}[ht]
\captionsetup{font=normalsize}
\renewcommand{\captionfont}{\linespread{1.6}\normalsize}
\centering
\caption{Flow rate and the corresponding Reynolds number and P\'{e}clet number in the experiments.}
\label{table 2}
\begin{tabular}{|c|c|c|}
\hline
\multirow{2}{*}{Flow Rate (mL/h)} & \multirow{2}{*}{Reynolds Number} & \multirow{2}{*}{P\'{e}clet Number} \\
 &  &  \\ \hline
15 & 1.37 & 458 \\ \hline
30 & 2.73 & 916 \\ \hline
60 & 5.47 & 1,832 \\ \hline
90 & 8.20 & 2,747 \\ \hline
120 & 10.94 & 3,663\\ \hline
150 & 13.67 & 4,579 \\ \hline
\end{tabular}
\end{table}

Seven spacers of different thickness from 90 $\mu m$ to 1,000 $\mu m$ were used to investigate the effect of channel height. The buoyancy and gravity effects of different channel heights can be estimated by a dimensionless number, Rayleigh number ($Ra$), as shown in Equation (\ref{Ra}) \cite{Zhang9253}. Table \ref{table 1} showed the corresponding Rayleigh number of each channel height.

\begin{equation}
\label{Ra}
Ra=\frac{\Delta \rho g (h/2)^3}{\mu D}
\end{equation}
where $\Delta \rho$ is the density difference between solution A and solution B, $g$ is the gravitational constant, $h$ is the channel height, $\mu$ is the viscosity of solution B, and $D$ is the diffusion coefficient.

\begin{table}[ht]
\captionsetup{font=normalsize}
\renewcommand{\captionfont}{\linespread{1.6}\normalsize}
\centering
\caption{Channel height and the corresponding Rayleigh number in the experiments.}
\label{table 1}
\begin{tabular}{|c|c|}
\hline
\multirow{2}{*}{Channel Height ($\mu m$)} & \multirow{2}{*}{Rayleigh Number} \\
 &  \\ \hline
90 & 219 \\ \hline
180 & 1,753 \\ \hline
260 & 5,283 \\ \hline
340 & 11,814 \\ \hline
430 & 23,899 \\ \hline
510 & 39,873 \\ \hline
\end{tabular}
\end{table}

To study the influence of chamber orientation, the device filled with solution A was placed horizontally and vertically (Figure \ref{chamber_5}(c)-(d)). Solution B was then injected at 30 $mL/h$.
Four elevated temperatures of 35 $^{\circ}$C, 50 $^{\circ}$C, 58 $^{\circ}$C, and 65 $^{\circ}$C were compared with the results of the ambient temperature. The densities, diffusion coefficients and viscosities at those temperatures are listed in Table~\ref{parameters}. All the solutions were immersed in the water bath for 30 $min$ before conducting the experiments to attain thermal equilibrium. The entire solvent exchange system, including the horizontally placed microchamber and pipes, were immersed into the water bath to keep the temperature constant during the experiments. The channel height was kept 1,000 $\mu m$ and the flow rate of solution B maintained a constant flow rate of 90 $mL/h$ for all the elevated temperatures.

\subsection{Observation and characterization of the asphaltene precipitates}
Total internal reflection fluorescence (TIRF) mode of a confocal microscope (Leica TCS SP8) was used to capture high-resolution images. A 488 $nm$ laser was used to excite the fluorophores of asphaltene. The spatial resolution was $\sim$ 200 $nm$ to show the details of asphaltene particles at the submicron scale. An optical microscope (Nikon ECLIPSE Ni) equipped with Nikon DS-FI3 camera was used to detect asphaltene particles at the micron scale. MATLAB (The MathWorks, Inc., US) and ImageJ (open source software) were used to analyze the images. The details of the process for image analysis can be found in our previous work \cite{meng2021microfluidic,MENG2021120584}. In brief, the main steps were to transform the microscopy images into binary images and then to obtain the information of particles such as perimeter and area. The particle size distribution smaller than one micron was obtained from the TIRF images, while the surface coverage and particle size distribution larger than one micron were obtained from the optical microscope images. 


\subsection{Description of the numerical model}
\textcolor{black}{In the present study, Level-Set (LS) method is coupled with the Transport of Diluted Species (TDS) physics adopted to track the interface between the two fluids and the concentration of the dissolved species at the oversaturation region.  }
\subsubsection{Transport of diluted species}
The transport of diluted species model solves the chemical species transport through diffusion and convection. This method resolves the mass conservation equation for oversaturation pulse, which can be written as: 
\begin{equation}
\label{eq:tdseq}
\frac{ \partial C_{i}}{ \partial t} +  \nabla.  J_{i}  +  \vec {U}. \nabla C_{i}=  0
\end{equation}
where $C_{i}$, $D_{i}$, $\vec{U}$, and $J_{i}$ are the concentration of species, diffusion coefficient, mass average velocity vector, and mass diffusive flux vector, respectively. The mass flux $J_{i}$ is further defined as $J_{i} = - D_{i} \nabla C_{i} $.

\subsubsection{Level-Set model}
A conservative level-set method \cite{sussman1994level} was combined with the transport of diluted species model to visualize the effect of gravity. In addition to the aforementioned equation for transport of diluted species, a set of coupled governing equations consisting continuity, momentum (incompressible Navier-Stokes equation), and a level-set equation was solved. 
\textit{Equation of continuity:}
\begin{equation}
\label{eq:mass_eqn}
  \nabla .   \vec{ U }  =0
\end{equation}
\textit{Equation of motion:} 
\begin{equation}
\label{eq:mom_eqn}
\rho \frac{ \partial ( \vec{ U })}{ \partial t} + \rho \nabla.( \vec{ U } \vec{ U }) = - \nabla P + \nabla.\mu (\nabla \vec {U} + \nabla { \vec {U} } ^{T}) + \rho g + \vec{ F}_{SF}
\end{equation}
where $\vec{U }$, $\rho $, $\mu$, $P$, and $\vec{ F}_{SF}$ are velocity, density, dynamic viscosity of fluid, pressure, and interfacial tension force, respectively.\\ \\
\textit{Equation of Level-Set (LS) Function:} \\
The interface in a level-set method is
captured by a conservative level set method (CLSM). The level function
changes smoothly across the interface from 0 to
1, when $\phi=0$, it indicates one fluid domain and $\phi=1$ in another fluid domain. The smooth interface between the phases is defined by the 0.5 isocontour of $\phi$. The following equation describes the convection of the reinitialized level set function \cite{mccaslin2014}.
\begin{equation}
\frac{\partial \phi}{\partial t}+\mathbf{u} \cdot \nabla \phi=\gamma \nabla \cdot\left(\epsilon \nabla \phi-\phi(1-\phi) \frac{\nabla \phi}{|\nabla \phi|}\right)
\end{equation}
where $\epsilon$ is a parameter to determine the thickness of the transition layer where $\phi$ goes smoothly from zero to one. It is typically of the same order as the size of the elements of the mesh. The  parameter $\gamma$ determines the amount of reinitialization of stabilization  of level set function \cite{COMSOL}. \textcolor{black}{The coupling of transport of diluted species interface with the level\textendash set physics adds an additional transport equation for computing the concentration of dissolved species “C”. The convective transport of “C” is based on the velocity field from the two\textendash phase flow, level\textendash set physics.}


\subsubsection{Computational parameters, implementation and model validation}

\begin{figure}[ht]
\centering
\includegraphics[width=1\columnwidth]{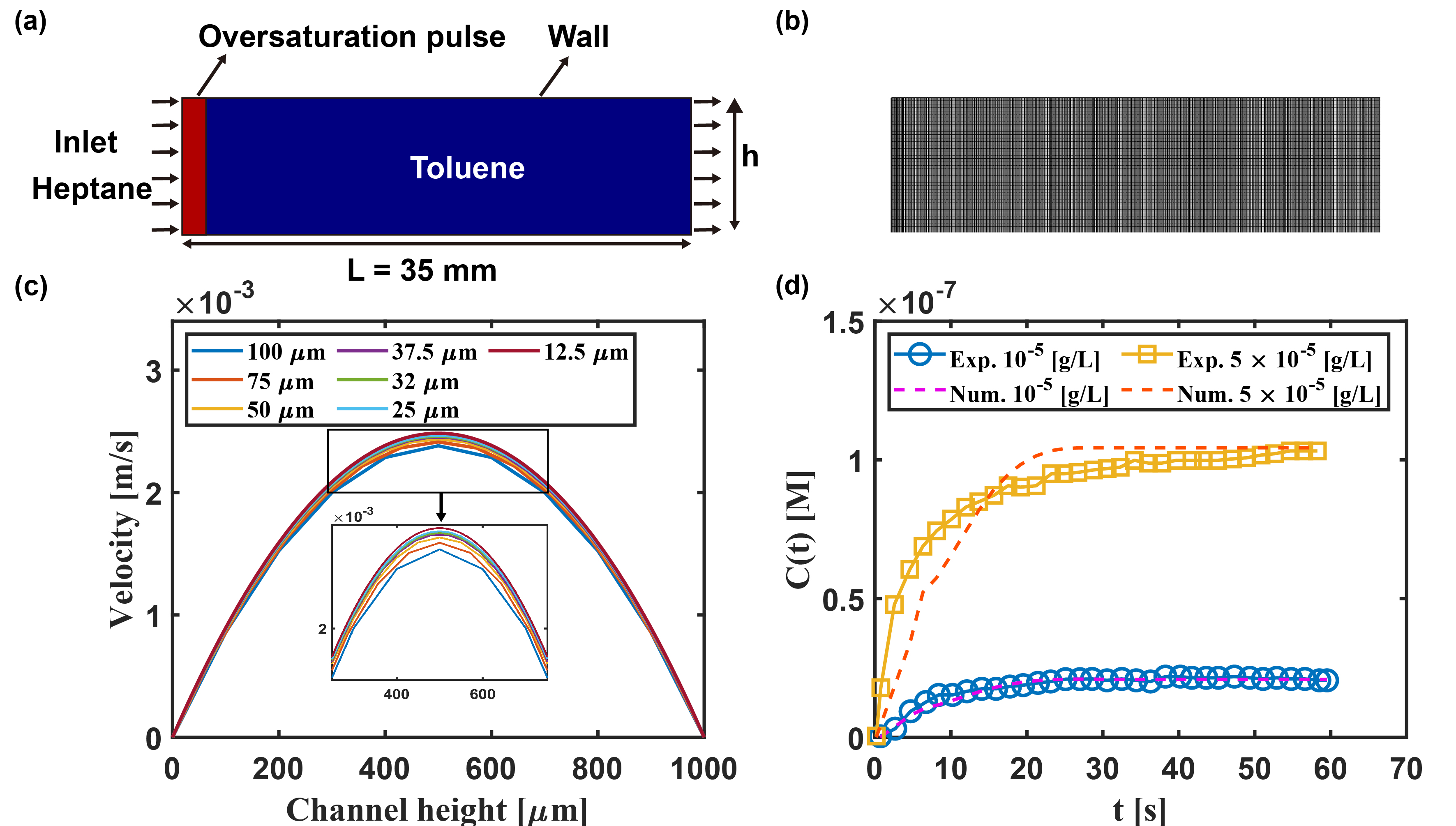}
\renewcommand{\captionfont}{\linespread{1.6}\normalsize}
\caption{(a) 2D schematic representation of the computational domain with imposed boundary conditions, and (b) computational gird. (c) Influence of grid numbers on the inlet velocity profiles along the height of the channel. (d) Model Validation with the experimental results reported in our previous work \cite{wei2022interfacial}. Blue circles and yellow squares represent the experimental data. Pink and red dashed lines represent the numerical results from our model for the same system.}
\label{CFD model}
\end{figure}

Systematic numerical investigations were performed to understand the mixing condition during solvent exchange process inside the microchamber. Two-dimensional transient simulations were carried out using a finite element method based solver COMSOL Multiphysics V. 6.0. (COMSOL AB, Stockholm, Sweden). \textcolor{black}{Asphaltene dissolved in toluene was considered as a fluid\textendash 1 and n\textendash heptane (solvent for asphaltene precipitation) was considered as fluid\textendash 2 in the computational domain. All the details of the fluid properties (i.e., density and viscosity) corresponding to experimental conditions are provided in the supporting information in Table S1. For different temperature conditions, toluene and n\textendash heptane properties were estimated using UNIFAC thermodynamic model with the software package Symmetry \cite{package}.} Influence of different microchannel geometries (i.e., height, see Table \ref{table 1}) and various flow rates (see Table \ref{table 2}) were examined . Unsteady state diffusion equation coupled with the velocity field was solved to obtain the evolution of the mixing front with space and time \cite{C6LC01555G}. The velocity inlet boundary condition was applied to the microchannel inlet, and the pressure outlet boundary condition was used at the outlet as shown in Figure \ref{CFD model}(a). 

\begin{table}[ht]
\captionsetup{font=normalsize}
\centering
\caption{Meshes in the grid independence study}
\renewcommand{\captionfont}{\linespread{1.6}\normalsize}
\label{grid}
\begin{tabular}{|c|c|c|c|c|c|}
\hline
Grid & \begin{tabular}[c]{@{}c@{}}Total \\ number of \\ elements\end{tabular} & \begin{tabular}[c]{@{}c@{}}Number\\ of boundary\\ elements\end{tabular} & \begin{tabular}[c]{@{}c@{}}Maximum\\ element\\ size (mm)\end{tabular} & \begin{tabular}[c]{@{}c@{}}Average\\ mesh quality\\ (skewness)\end{tabular} & \begin{tabular}[c]{@{}c@{}}Difference (\%)\\ w.r.t G7\end{tabular} \\ \hline
G1 & 3,500 & 730 & 0.10000 & 1.0 & 4.06 \\ \hline
G2 & 6,552 & 978 & 0.0750 & 1.0 & 2.81 \\ \hline
G3 & 14,000 & 1,460 & 0.0500 & 1.0 & 1.76 \\ \hline
G4 & 25,218 & 1,949 & 0.0375 & 1.0 & 1.34 \\ \hline
G5 & 35,008 & 2,248 & 0.0320 & 1.0 & 0.95 \\ \hline
G6 & 56,000 & 2,920 & 0.0250 & 1.0 & 0.51 \\ \hline
G7 & 224,000 & 5,840 & 0.0125 & 1.0 & - \\ \hline
\end{tabular}
\end{table}

No-slip boundary condition was imposed at the top and bottom walls and inflow with zero concentration at the inlet. At $t=0$, the oversaturation pulse in a narrow region (i.e., 100 $\mu m$ thick) was initialized vertically along the inlet with a uniform concentration. The oversaturation pulse was convected and diffused simultaneously inside the microchannel, once the flow is initialized. The computational domain was meshed with quadrilateral mesh elements (Figure \ref{CFD model}(b)). Grid independence study was initially performed using different element sizes ranging from 0.10 to 0.0125 $mm$. Mesh quality and number of elements details are reported in Table \ref{grid}. The velocity profiles were compared for different meshes as shown in Figure \ref{CFD model}(c), and an optimum mesh element size of 0.025 $mm$ was considered for the numerical investigations. Different sets of simulations were performed in a high performance computing (HPC) facility for the same flow conditions ($Pe$, $Ra$, $Re$, $h$, and $Q$) as used in the experiments . For each simulation, 48 CPUs were used to run the 30 $s$ of flow time.




At first, the developed CFD model was validated with the relevant experimental data from our previous study \cite{wei2022interfacial}. The length, width and the height of the considered microchamber are 45 $mm$, 13 $mm$, and 0.3 $mm$ respectively, which are within the range of our current microchamber dimensions. The flow rate at the inlet was set to 50 $mL/h$ and other fluid properties were kept similar to our previous study \cite{wei2022interfacial}. Figure \ref{CFD model}(d) shows a comparison of the analyte concentration as a function of time. The CFD model predictions were found to be in good agreement with experimental data. \textcolor{black}{Therefore, the developed CFD model was capable to predicted underlying phenomena of solvent exchange process.}



\section{Results and Discussion}

In the solvent exchange process, solution B gradually displaces solution A. Simultaneously, asphaltene from solution A diffuses to solution B, and n-heptane diffuses towards solution A. As the diffusion process advances, The S/B ratio in the mixing zone gradually increases beyond the onset, until the concentration of asphaltene in the mixing front is oversaturated. At t = 0, the asphaltene in solution A is undersaturated. As time progresses, the asphaltene becomes oversaturated in the mixing region, which triggers the formation of asphaltene precipitate. Once the solution A is completely displaced by the solution B, the formation of asphaltene precipitate stops due to lack of asphaltene. The duration of the oversaturation pulse is independent of the flow velocity due to the no-slip boundary and depends on the diffusion of chemicals from the mixing front to the substrate \cite{Zhang9253}. Therefore, the new phase formation on the substrate is determined by the profile of the mixing front.

\subsection{Effect of the flow rate}

\begin{figure}[ht]
\centering
\includegraphics[width=1\columnwidth]{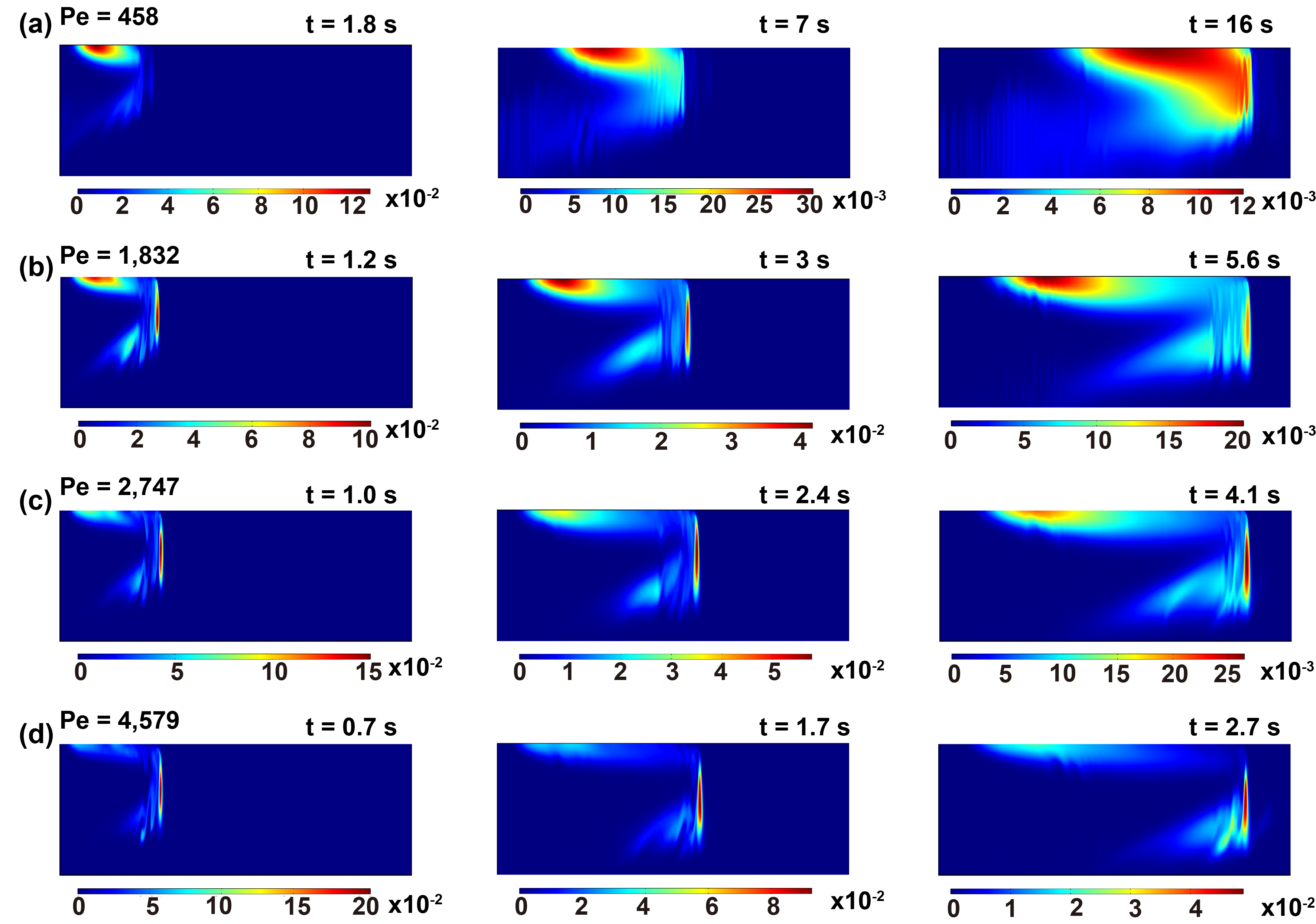}
\renewcommand{\captionfont}{\linespread{1.6}\normalsize}
\caption{COMSOL simulation for the profile of the oversaturation pulse at (a) $Pe$ = 458, (b) $Pe$ = 1,832, (c) $Pe$ = 2,747, and (d) $Pe$ = 4,579.
}
\label{COMSOL_Pe}
\end{figure}

Simulation results show the oversaturation pulse profile for different flow rates in Figure~\ref{COMSOL_Pe}. 
The non-blue area is the mixing front, and the color scale bar indicates the corresponding concentration of oversaturation pulse in the system. At $Pe$ of 458, the oversaturation pulse passes through the entire chamber in 16 $s$. At the highest $Pe$ of 4,578, the duration of the pulse is approximately 2.7 $s$. Although the time for the precipitation is considerably shorter at larger $Pe$, the oversaturation levels near the bottom wall are similar through the entire process at all $Pe$, as exhibited by the consistent blue zone at the bottom in all snapshots. Therefore, diffusion of asphaltene and n-heptane towards the substrate at different $Pe$ should be comparable because the concentration gradient is identical, resulting in a similar $SC$ of asphaltene on the bottom wall. At any given moment, the level of the oversaturation on the top plate is substantially higher at lower $Pe$, which may lead to more precipitates on the top surface.

\begin{figure}[ht]
\centering
\includegraphics[width=1\columnwidth]{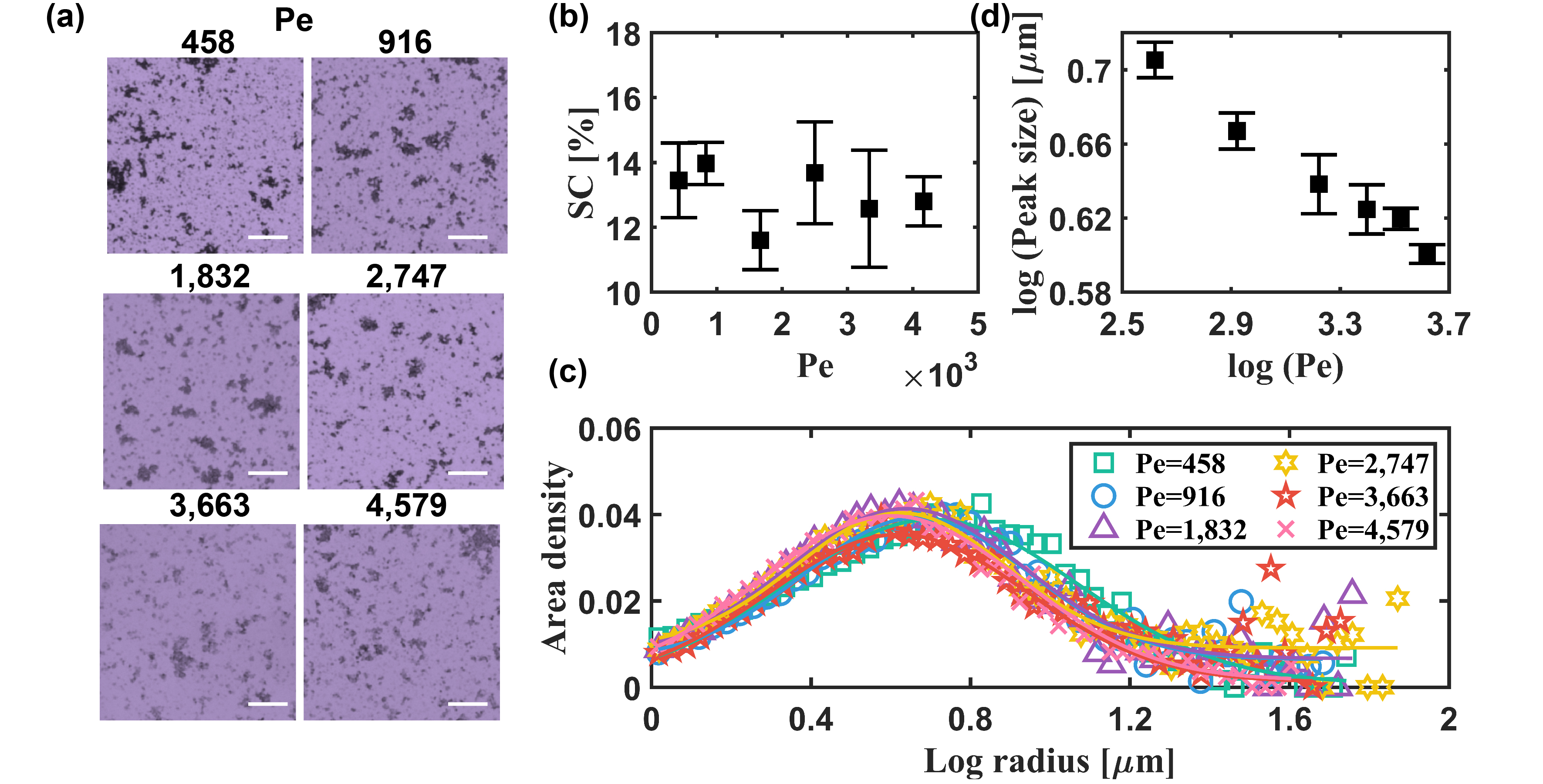}
\renewcommand{\captionfont}{\linespread{1.6}\normalsize}
\caption{(a) Optical images of asphaltene particle at the final state of different P\'{e}clet numbers. The length of the scale bar is 20 $\mu m$. The images are false-colored. (b) $SC$ of asphaltene particles at the final state as a function of P\'{e}clet number. (c) Size distribution of asphaltene particles for different $Pe$. The solid lines are the Gaussian fitting for the results. (d) The corresponding peak size of Gaussian fitting of (c) vs. P\'{e}clet number.}
\label{flow rate results}
\end{figure}

Figure \ref{flow rate results}(a) shows the optical images of asphaltene precipitation in the solvent exchange with the flow rate of solution B ranging from 15 $mL/h$ to 150 $mL/h$ (i.e., $Pe$ from 458 to 4,579). 
The black dots are asphaltene particles formed on the substrate placed on the bottom wall. It is worth mentioning that only the area with asphaltene particles is considered. The areas of asphaltene layers, as shown in Figure \ref{1000}(b), are omitted when analyzing surface coverage ($SC$) and the size distribution of asphaltene particles because these large deposits may be agglomerates formed in the bulk. Figure \ref{flow rate results}(b) shows $SC$ of the asphaltene particles divided by the total area, which is approximately independent of $Pe$ and is always $\sim$ 14\%.

Figure \ref{flow rate results}(c) shows the size distribution of asphaltene particles. The distribution is plotted based on area density instead of the number density \cite{maqbool2011modelling} to compare with the conventional study in a bulk system. Interestingly, the Gaussian distribution can fit the area density-based size distribution for each $Pe$. This Gaussian distribution is consistent with the presented results in a conventional mixing system \cite{maqbool2011modelling}. The peaks of the size distribution in Figure \ref{flow rate results}(c) shift to the left with the increase of $Pe$, indicating the size of asphaltene particles decreases with the increase of $Pe$ (Figure \ref{flow rate results}(d)). 

\begin{figure}[ht]
\centering
\includegraphics[width=1\columnwidth]{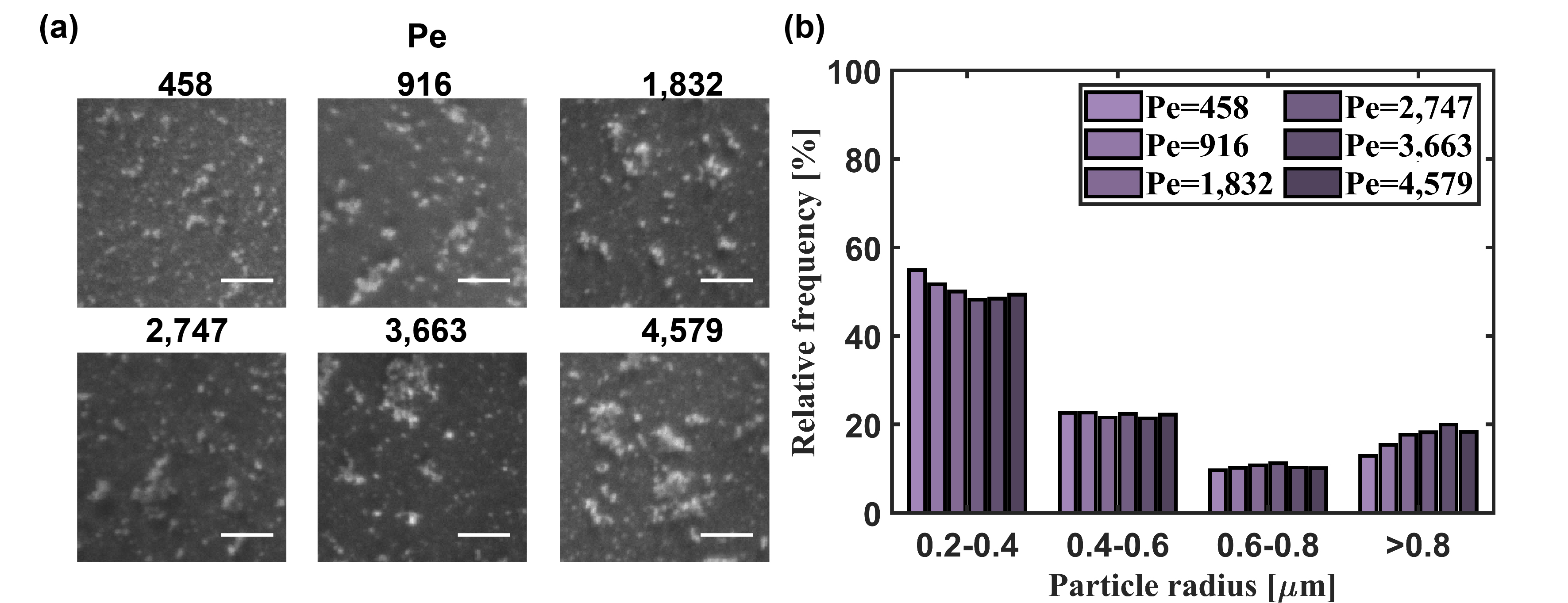}
\renewcommand{\captionfont}{\linespread{1.6}\normalsize}
\caption{(a) TIRF images of asphaltene particles at the final state of different $Pe$. The length of the scale bar is 5 $\mu m$. (b) Relative frequency of size distribution of asphaltene particles at different flow rates.
}
\label{flow rate TIRF}
\end{figure}

Figure \ref{flow rate TIRF}(a) shows the TIRF images of asphaltene particles at different $Pe$. 
The white speckles are the asphaltene particles due to the natural fluorescence property of asphaltene under 488 $nm$ laser excitation in air. Number density-based relative frequency is introduced to analyze the size distribution quantitatively. The relative frequency of each size bin is calculated by the number of particles of the corresponding size divided by the total number of particles. For primary submicron particles (PSMPs), which are defined as the radius from 200 to 400 $nm$ \cite{meng2021primary}, the relative frequency denotes the degree of aggregation at the early stage. A high relative frequency of PSMPs ($R_p$) represents the low possibility of PSMPs aggregating into larger particles after the precipitation. From the quantitative examination, as shown in Figure \ref{flow rate TIRF}(b), it can be observed that $R_p$ decreases with the increase of $Pe$. \textcolor{black}{$R_p$ is calculated by:}

\begin{equation}
\color{black}
\label{Rp}
R_p = \frac{N_{psmp}}{N_{psmp}+N_{smp}}
\end{equation}
\textcolor{black}{where $N_{psmp}$ and $N_{smp}$ are the quantity of primary submicron particles and quantity of submicron particles, respectively.}

Interestingly, the size of asphaltene particles correlates with the flow rate because of the mass transfer process of asphaltene from the oversaturation pulse to the substrate. The size of asphaltene particles decreases with $Pe$ at micron-scale (Figure \ref{flow rate results}(c)) but increases at submicron scale (Figure \ref{flow rate TIRF}(b)). The particle aggregation and fragmentation process are discussed to examine the reasons for the opposite trend of size distribution at submicron and micron scales. Based on the Smoluchowshi aggregation model, the aggregation is:

\begin{equation}
\label{Sm_5}
\frac{dn_k}{dt}=\frac{1}{2} \sum_{i+j=k} K_{i,j} n_i n_j - n_k \sum_{i \ge 1} K_{i,k} n_i
\end{equation}
where $n_i$, $n_j$, and $n_k$ is the number density change of particle of size $i$, $j$, and $k$, respectively. $t$ is time, and $K_{i,j}$ is aggregation kernel, which is calculated by $K_{i,j} = \alpha_{i,j} \beta_{i,j}$, where $\alpha_{i,j}$ is the collision frequency and $\beta_{i,j}$ is the collision efficiency of particles. $\beta_{i,j}$ is:

\begin{equation}
\label{beta_5}
\beta_{i,j} \propto exp [- \frac{1}{k_B T (\delta_{asp} - \delta_{sol})^2}]
\end{equation}
where $k_B$ is Boltzmann constant, $\delta_{asp}$ and $\delta_{sol}$ are Hildebrand solubility parameters of asphaltene and the paraffinic solvent, respectively. $\alpha_{i,j}$ is driven by two factors in our system, including Brownian motion and shear force \cite{nassar2015development,Ali2008Dynamic,thomas1999flocculation,rahmani2003characterization,rahmani2004evolution}:  

\begin{equation}
\label{alpha}
\alpha_{i,j} = \frac{2}{3} \frac{RT}{\mu} \frac{(d_i + d_j)^2}{d_i.d_j} + \frac{G}{6}(d_i + d_j)^3
\end{equation}
where $R$ is the ideal gas constant. $T$ is temperature. $\mu$ is viscosity. $d_i$, and $d_j$ are the diameters of particles with sizes $i$ and $j$, respectively, $G$ is the shear rate.

The oversaturation level near the bottom wall is similar for all the $Pe$ (Figure \ref{COMSOL_Pe}). Regardless, the velocity gradient between the maximum point of the flow and the minimum point adjacent to the substrate increases with $Pe$. The increased shear force leads to the increase of aggregation of PSMPs at an early stage (Equation (\ref{alpha})), ultimately resulting in lowering of $R_p$ (Figure \ref{flow rate TIRF}(b)).

However, shear force brings an increase in the aggregation and the fragmentation of aggregates. The mechanism of fragmentation induced by shear rate is due to different pressures gradient on the opposite sides of the aggregates \cite{parker1972floc}. The fragmentation rate ($B_i$) increases with the particle volume ($V_i$) and the shear rate ($B_i = k G^y V_i^{1/3}$), where $k$ is a proportionality fitting constant, $y$ is a constant inversely proportional to the interaction strength between particles \cite{rahmani2003characterization,kapur1972self,boadway1978dynamics,pandya1982floc}. Therefore, when the asphaltene particles are large enough (i.e., at the micron scale), the fragmentation yielded by shear force prevails, resulting in the decrease of particle size with the increase of shear rate (Figure \ref{flow rate results}(d)). \textcolor{black}{It is worth noting that there is no maximum critical size for the asphaltene particles in the solvent exchange process because the chemical composition varies with time.}

\subsection{Effect of the microchamber height}

\begin{figure}[ht]
\centering
\includegraphics[width=1\columnwidth]{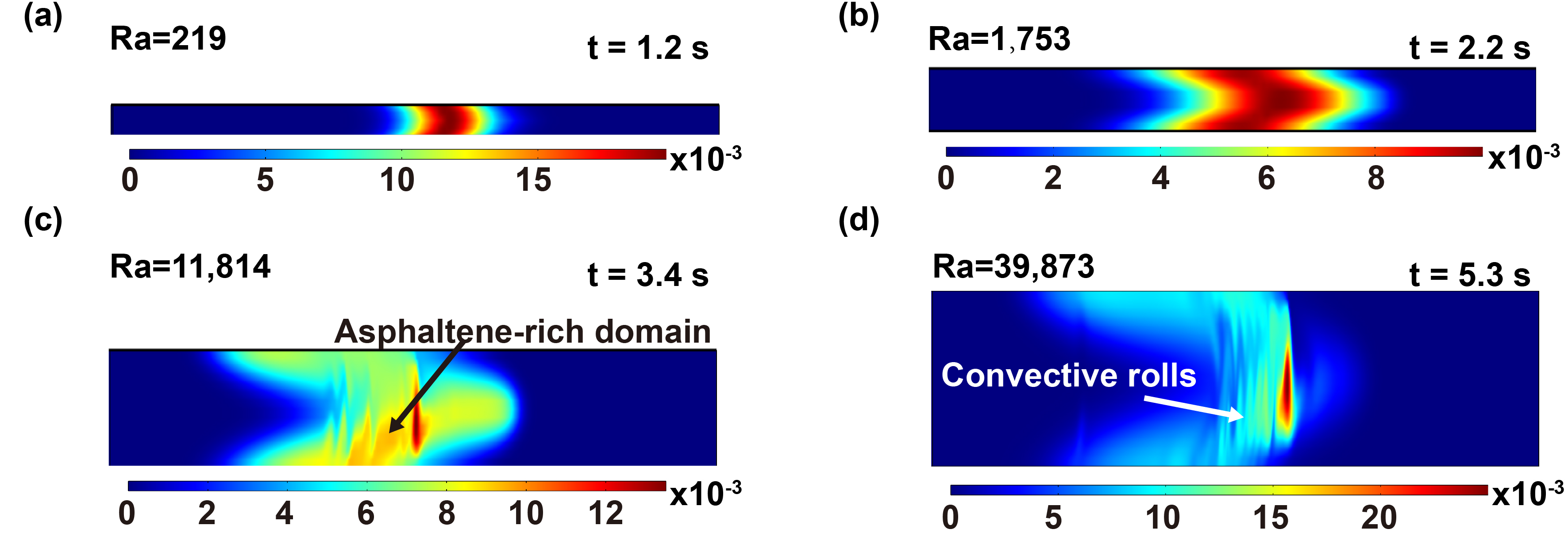}
\renewcommand{\captionfont}{\linespread{1.6}\normalsize}
\caption{COMSOL simulation for the profile of the oversaturation pulse at (a) $Ra$ = 219, (b) $Ra$ = 1,753, (c) $Ra$ = 11,814, and (d) $Ra$ = 39,873. 
}
\label{COMSOL channel height}
\end{figure}

A dimensionless number (Rayleigh number, $Ra$) is used to quantitatively analyze the impact of channel height on $SC$ (Table \ref{table 1}). The oversaturation levels near the bottom wall are always high, as exhibited by the red color in Figure \ref{COMSOL channel height}.  At $Ra$ of 219, the oversaturation pulse passes through the chamber in 2.1 $s$. At the highest $Ra$ of 39,873, the duration of the pulse is approximately 9.3 $s$. The increased duration of the oversaturation pulse provides a longer time for asphaltene precipitation and aggregation. Therefore, mass transfer of asphaltene from the oversaturation pulse to the substrate should increase with the increase of the $Ra$, resulting in a higher $SC$ and larger size of asphaltene particles.

\begin{figure}[ht]
\centering
\includegraphics[width=1\columnwidth]{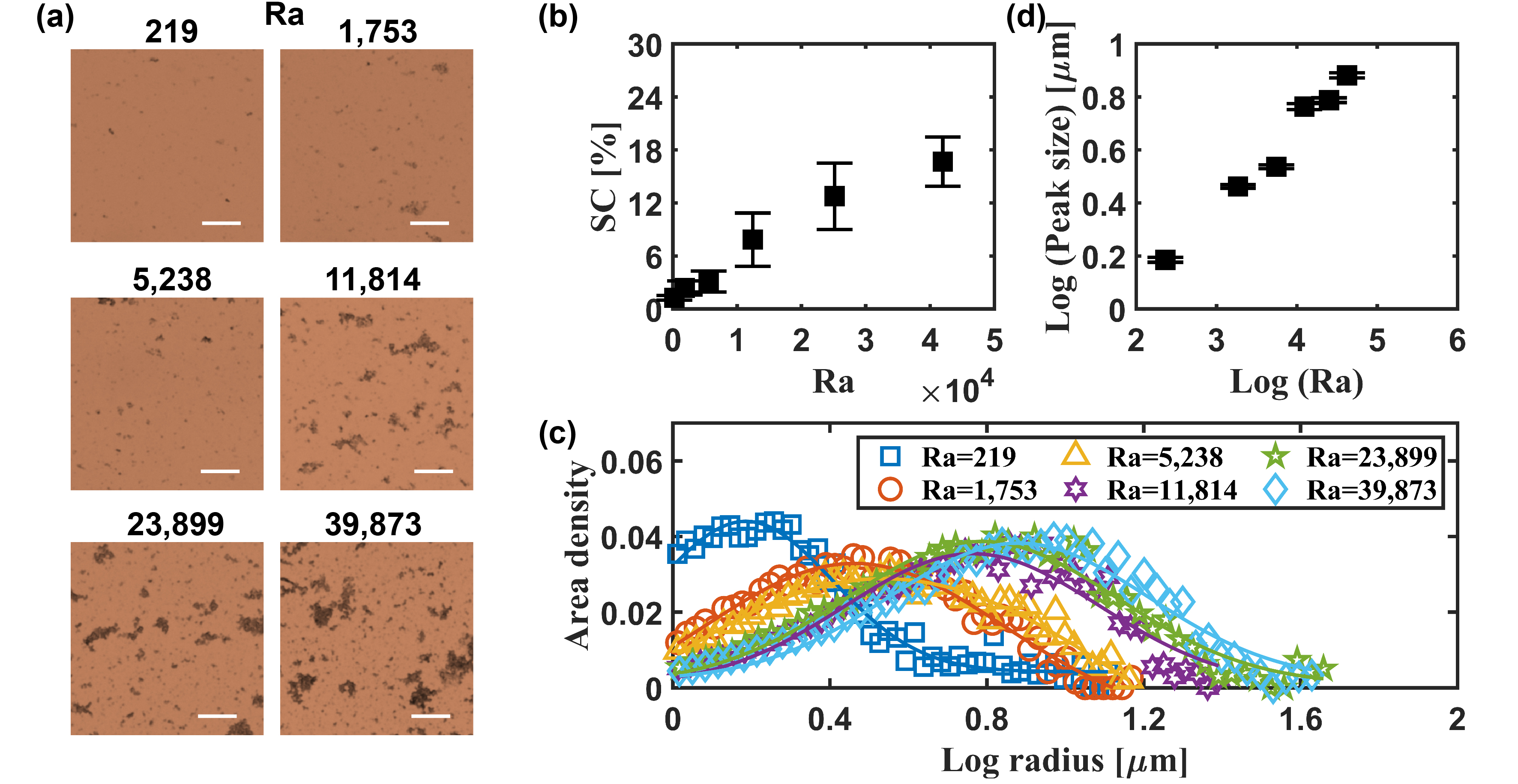}
\renewcommand{\captionfont}{\linespread{1.6}\normalsize}
\caption{(a) Optical images of asphaltene particles at the final state for different $Ra$. The length of the scale bar is 20 $\mu m$. The images are false-colored. (b) $SC$ of asphaltene at the final state as a function of Rayleigh number. (c) Size distribution of asphaltene particles for different channel heights. The scatters are the experimental data and the solid lines are the Gaussian fitting for the results. (d) The corresponding peak size of Gaussian fitting of (c) vs. channel height. 
}
\label{channel height results}
\end{figure}

Figure \ref{channel height results}(a) shows the optical microscope images of asphaltene precipitates in channel heights from 90 $\mu m$ to 510 $\mu m$ under the same $Pe$. 
It is found that more asphaltene particles are formed with the increase of the channel height. Meanwhile, the size of asphaltene particles increases with the increase of the channel height. $SC$ increases with $Ra$ (Figure \ref{channel height results}(b)). The relationship is almost linear. However, the linear relationship is not held when the channel height is 1,000 $\mu m$ as shown in Figure \ref{1000}(a). At 1,000 $\mu m$, many asphaltene deposits do not exist as particles but form a layer of sediments, as shown in Figure \ref{1000}(b). \textcolor{black}{Therefore, $SC$ increases as the channel height increases and eventually reaches a maximum value.} The formation of a larger $SC$ and a layer of sediments of asphaltene are in line with the numerical findings. 
Figure \ref{channel height results}(c) shows the size distribution of asphaltene particles as a function of $Ra$. The size distribution for each $Ra$ can be fitted by Gaussian distribution. The Gaussian curve peaks increased from $\sim$ 0.2 to $\sim$ 0.9 with $Ra$ increasing from 219 to 39,873. It indicates that the mean particle radius increases from 1.6 $\mu m$ to 7.9 $\mu m$ as shown in Figure \ref{channel height results}(d). Particle size increases with $Ra$ because better solution mixing leads to an increase in particle collision efficiency $\beta$ in Equation (\ref{beta_5}), resulting in larger particles. These observations are also consistent with our numerical model predictions.



\begin{figure}[ht]
\centering
\includegraphics[width=1\columnwidth]{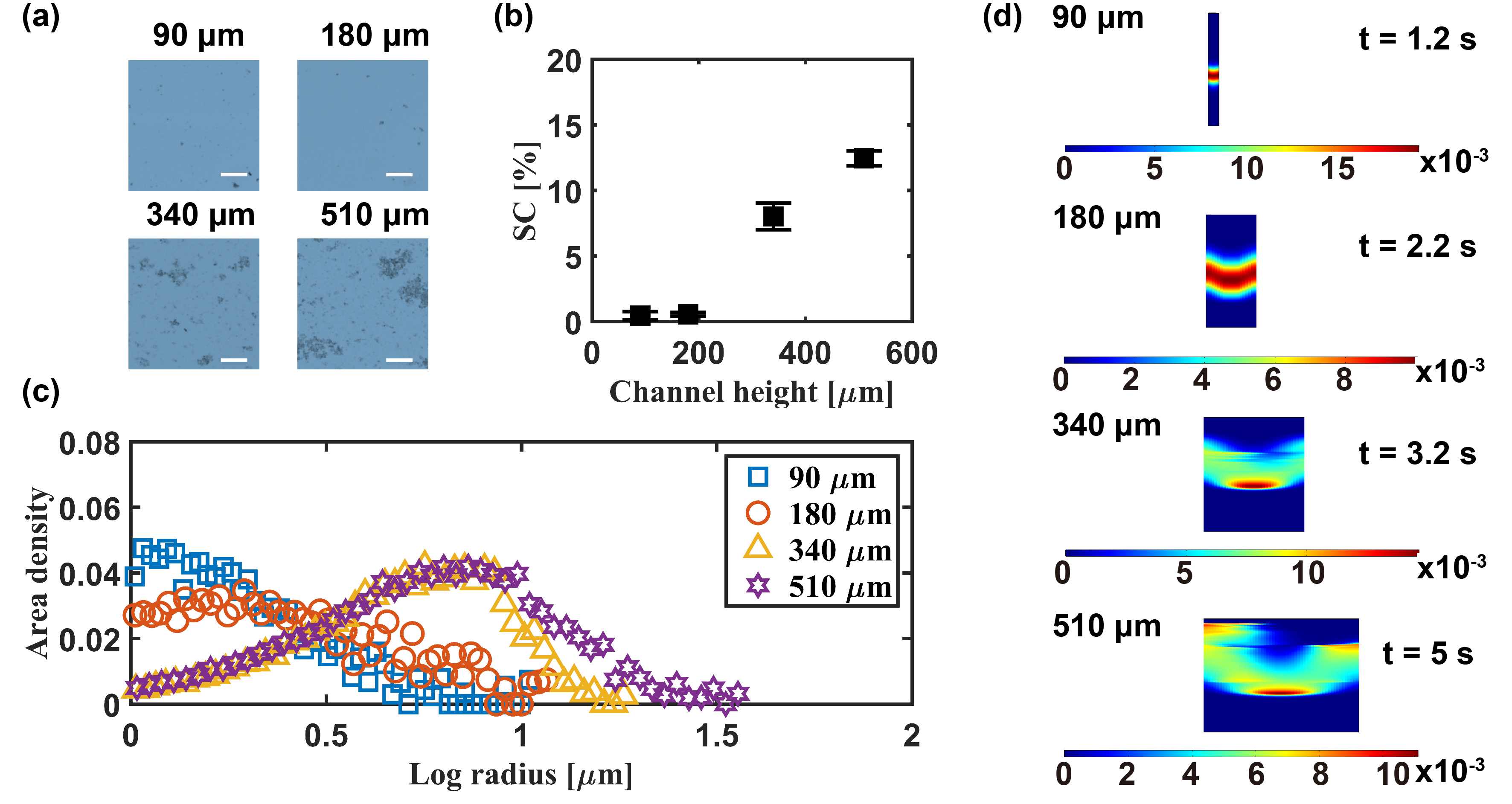}
\renewcommand{\captionfont}{\linespread{1.6}\normalsize}
\caption{(a) Optical images of asphaltene particles at the final state of different channel heights. The length of the scale bar is 20 $\mu m$. The images are false-colored. (b) Surface coverage of asphaltene at the final state as a function of channel heights. (c) Size distribution of asphaltene particles for different temperatures. The solid lines are the Gaussian fitting for the results. (d) COMSOL simulation for the profile of the oversaturation pulse at different channel heights.
}
\label{vertical placed}
\end{figure}

Our simulations for the vertically placed devices also demonstrate that the duration of the oversaturation pulse increases with the increase of the channel height, as shown in Figure \ref{vertical placed}(d), which is consistent with the horizontally placed devices. Therefore, the $SC$ increases with the channel height and the size distribution shifts to the right with the increase of channel height (Figure \ref{vertical placed}(a)-(c)). 

\begin{figure}[ht]
\centering
\includegraphics[width=1\columnwidth]{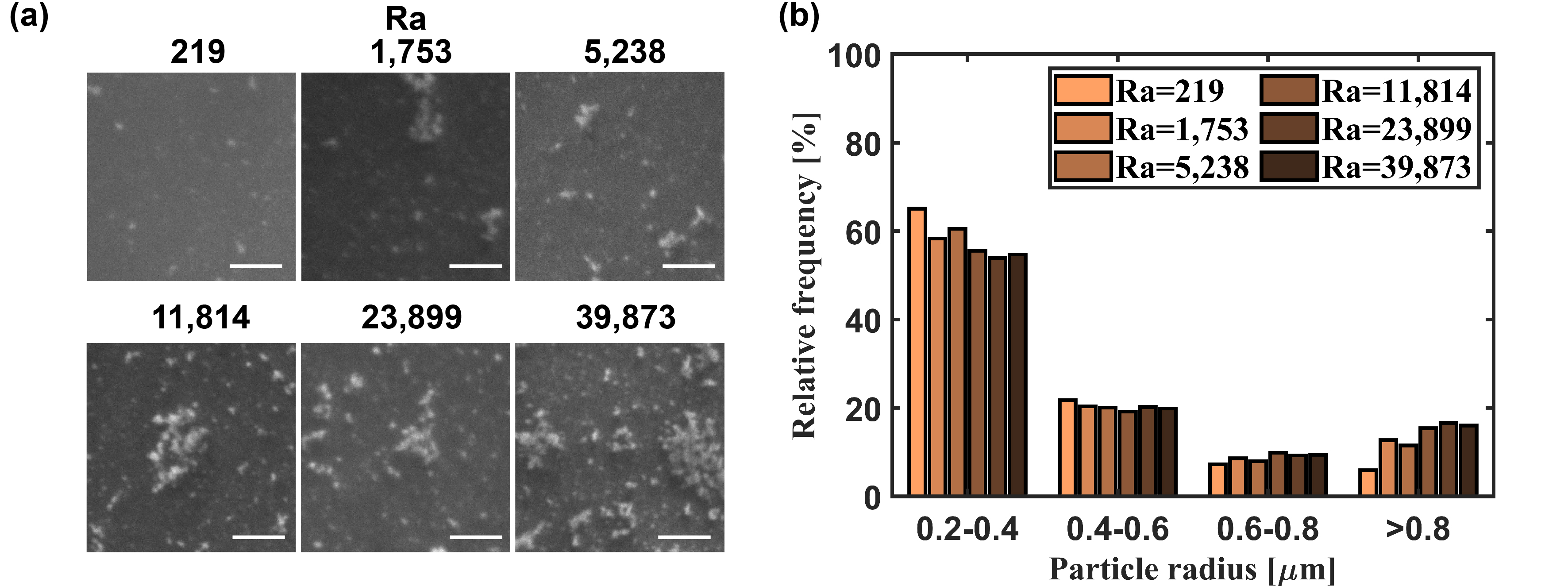}
\renewcommand{\captionfont}{\linespread{1.6}\normalsize}
\caption{(a) TIRF images of asphaltene particles at the final state for different channel heights. The length of the scale bar is 5 $\mu m$. (b) Relative frequency of the size distribution of asphaltene particles of different channel height.}
\label{channel height TIRF}
\end{figure}

TIRF images of the final state of asphaltene particles are captured to show the details of asphaltene particles at the submicron scale, and is shown in Figure \ref{channel height TIRF}(a). 
It can be seen from Figure \ref{channel height TIRF}(b) that $R_p$ decreases from $\sim$ 65\% to $\sim$ 52\% with the increase of $Ra$ from 219 to 39,873, reflecting the aggregation of particles is more intensified in the channel with a more significant height. Correspondingly, the proportion of large particles, especially those larger than 0.8 $\mu m$, increases with the increase of the channel height. This observation at the submicron scale is consistent with that at the micron scale. 


\subsection{Effect of the microchamber orientation}

In the device with the same channel height, the oversaturation pulse takes the same time to pass through the channel in both horizontal and vertical devices. However, the mixing dynamics are different. When the channel height is 510 $\mu m$ ($Ra$ is 39,873 for the horizontally placed device), simulation results showed the formation of convective rolls (Figure \ref{COMSOL channel height}(d)) in the horizontally placed device, leading to better mixing between solutions A and B. The final maximum oversaturation level for the horizontally placed device is approximately 0.02, while the one for the vertically placed device is approximately 0.01. In addition, the oversaturation area is not in the middle but shifts downwards and thus closer to the bottom wall because the asphaltene is heavier compared with n-heptane and toluene, an asphaltene-rich domain forms at the tail of the profile of the oversaturation pulse (Figure \ref{COMSOL channel height}(c)). However, the asphaltene-rich domain does not appear in the vertically placed cases. The loss of convective rolls and asphaltene-rich domain in the vertically placed devices results in reduced mass transfer compared to horizontally placed devices. Therefore, diffusion of asphaltene and n-heptane towards the substrate of the vertically placed device should be smaller than the horizontally placed device, resulting in the decrease of $SC$ and left-shift of the particle size distribution. 

Figure \ref{vertical}(a) shows the comparison of optical images of asphaltene precipitates between horizontally and vertically placed devices. The horizontally placed device has a higher $SC$ of asphaltene particles than the vertically placed device, especially for 510 $\mu m$ (Figure \ref{vertical}(b)). $SC$ increases from $\sim$ 12\% to $\sim$ 18\%, revealing more asphaltene deposits on the substrate, which is in line with the numerical results.

The numerical results of varying P\'{e}clet number displays stack-like extreme convective rolls and very large asphaltene-rich region (Figure \ref{COMSOL_1000}). Video S1 (see supplementary information) shows a backflow after the mixing front reaches the outlet. Therefore, the $SC$ may increase sharply when the channel height is 1,000 $\mu m$. The formation of the layer of sediments in the 1,000 $\mu m$ channel height group is consistent with our numerical results in Figure \ref{1000}(b). Due to the intense mixing of the model oil and solvents, the asphaltene precipitation at 1,000 $\mu m$ may be closer to that in the bulk phase mixing system in the conventional study \cite{xu2017asphaltene}.

Orientation direction can also affect the size distribution of the asphaltene particles deposited on the substrate. Figure \ref{vertical}(d)-(f) reveals the rightwards shift of particle size in the horizontally placed device, meaning the asphaltene particles are relatively large in horizontally placed devices compared with the vertically placed ones. However, the shift of size distribution is not observed in the 90 $\mu m$ microchamber (Figure \ref{vertical}(c)). This is due to $Ra$ is smaller than the critical value (1,708) when the channel height is 90 $\mu m$ \cite{Zhang9253}. The convective rolls and asphaltene-rich domain are not observed in the horizontally placed device.

In the solvent exchange process, particles may form on the substrate and in the bulk of the liquid simultaneously. The particles formed in the bulk may settle down onto the substrate under gravity in the horizontally placed device. This is mainly due to the competition between gravity and Brownian motion. Brownian motion of the particles ($\overline{X}_{Brownian}$) is \cite{allen1990particle}:

\begin{equation}
\color{black}
\label{Brownian}
\overline{X}_{Brownian}^2 = \frac{4RTK_mt}{3\pi^2 \mu Nd}
\end{equation}
where $R$ is the ideal gas constant, $T$ is temperature, $K_m$ is the correction for discontinuity of fluid, which is 1 for liquid. $t$ is time, which is calculated by dividing the volume of the microchamber by the flow rate, $\mu$ is the viscosity of the solvent, $N$ is Avogadro constant, \textcolor{black}{$d$ is the diameter of particles}. The Stokes settling motion of a particle is \cite{allen1990particle}:

\begin{equation}
\color{black}
\label{stokes}
\overline{X}_{Settling} = \frac{(\rho_{asp}-\rho_{sol}) g d^2 t}{18 \mu}
\end{equation}
where $\rho_{asp}$ is the density of asphaltene and $\rho_{sol}$ is the density of the solvent. For each channel height, only the particles with a diameter larger than 8 $\mu m$ (logarithmic radius is 0.6) can settle under gravity. However, the rightwards shift is observed of particles smaller than 8 $\mu m$ as shown in Figure \ref{vertical}(d). In addition, particles larger than 8 $\mu m$ in Figure \ref{vertical}(c) of 90 $\mu m$ device do not show a rightwards shift. Therefore, gravity-induced settlement is not the reason for the rightwards shift in horizontally placed devices. The increase of particle size is also due to the enhancement of solution mixing in the horizontally placed device.


\subsection{Effect of the temperature}

\begin{figure}[ht]
\centering
\includegraphics[width=\columnwidth]{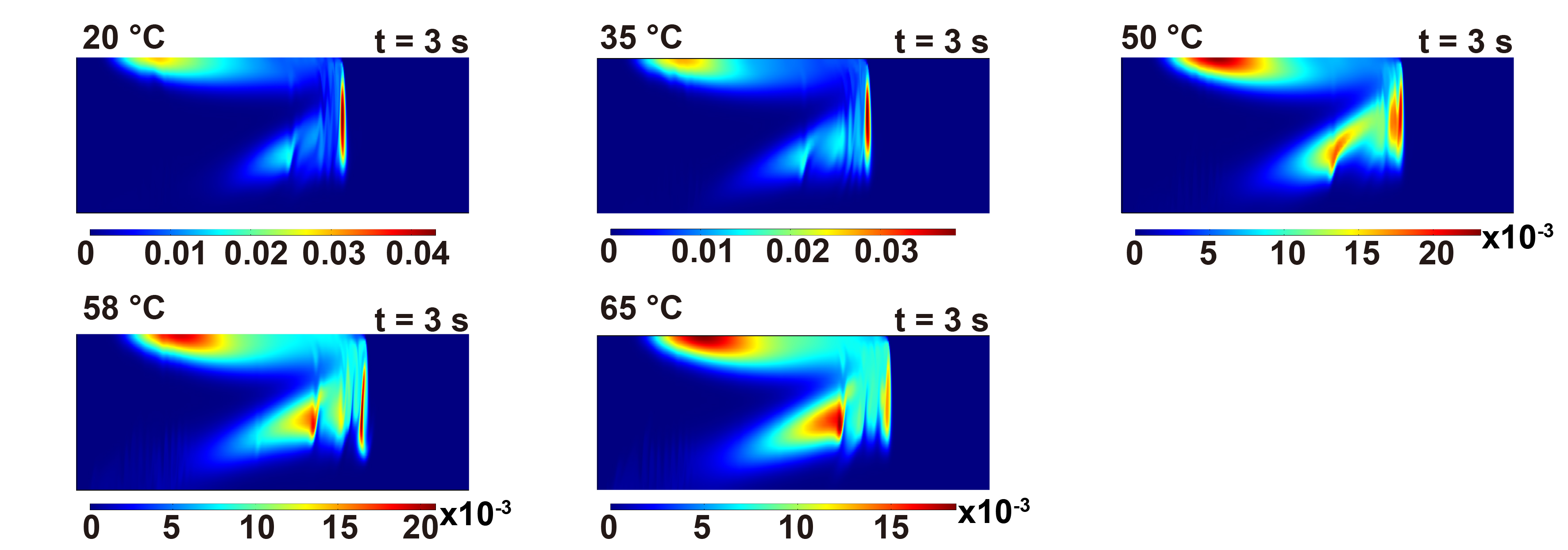}
\renewcommand{\captionfont}{\linespread{1.6}\normalsize}
\caption{COMSOL simulation results of the oversaturation profile at 20 $^\circ C$, 35 $^\circ C$, 50 $^\circ C$, 58 $^\circ C$, and 65 $^\circ C$.}
\label{COMSOL_temp}
\end{figure}

The density difference between solutions A and B does not change significantly with the change in temperature (Table \ref{parameters}). The duration for the oversaturation pulse to completely pass the entire channel is approximately 4 $s$ for all the examined temperatures. Figure \ref{COMSOL_temp} shows the profile of the oversaturation pulse near the bottom wall does not change much with the increase in temperature, as indicated by the consistent blue zone. 
In addition, from the experiments performed in the 1,000 $\mu m$ microchamber, it was evident that there \textcolor{black}{was} sufficient mixing in all of the temperatures due to the presence of convective rolls, as shown in Figure \ref{COMSOL_1000}. Therefore, mixing at all the temperatures we studied should be adequate. 



\begin{figure}[ht]
\centering
\includegraphics[width=1\columnwidth]{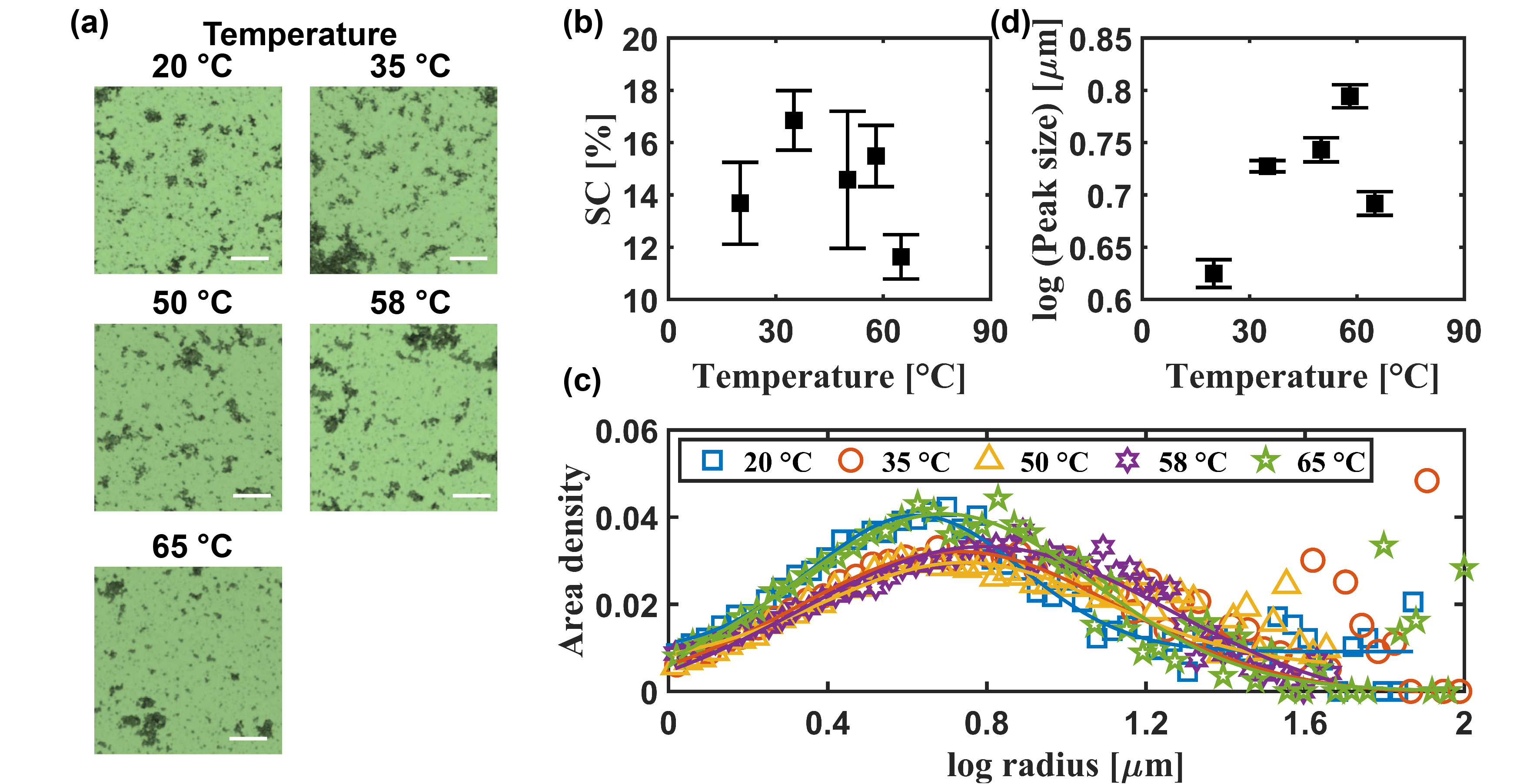}
\renewcommand{\captionfont}{\linespread{1.6}\normalsize}
\caption{(a) Optical images of asphaltene particles at the final state of different temperatures. The length of the scale bar is 20 $\mu m$. The images are false-colored. (b) Surface coverage of asphaltene at the final state as a function of temperature. (c) Size distribution of asphaltene particles for different temperatures. The solid lines are the Gaussian fitting for the results. (d) The corresponding peak size of Gaussian fitting of (c) vs. temperature.
}
\label{temp results}
\end{figure}

Figure \ref{temp results}(a) shows the optical images of asphaltene particles at different temperatures. 
$SC$ from 20 $^\circ C$ to 58 $^\circ C$ are similar but $SC$ of 65 $^\circ C$ is much smaller than the other groups. The size distribution shows a rightwards shift trend with the increase of temperature from 20 to 58 $^\circ C$. But the size distribution shifts to left from 58 to 65 $^\circ C$ (Figure \ref{temp results}(b)-(d)). The experiment results are mainly due to the thermodynamics influence. Asphaltene is a mixture of different fractions, and the increase in temperature resulted in some asphaltene fractions becoming soluble in asphaltene \cite{acevedo2010investigation}. \textcolor{black}{The reason may be the increase of steric repulsion between asphaltene particles caused by the increase in temperature \cite{wang2010interaction,wang2009colloidal}.} These asphaltene fractions do not precipitate, resulting in a decrease of $SC$ at 65 $^\circ C$. This observation is consistent with the previous study in a conventional bulk system \cite{xu2017asphaltene}.

\begin{figure}[ht]
\centering
\includegraphics[width=1\columnwidth]{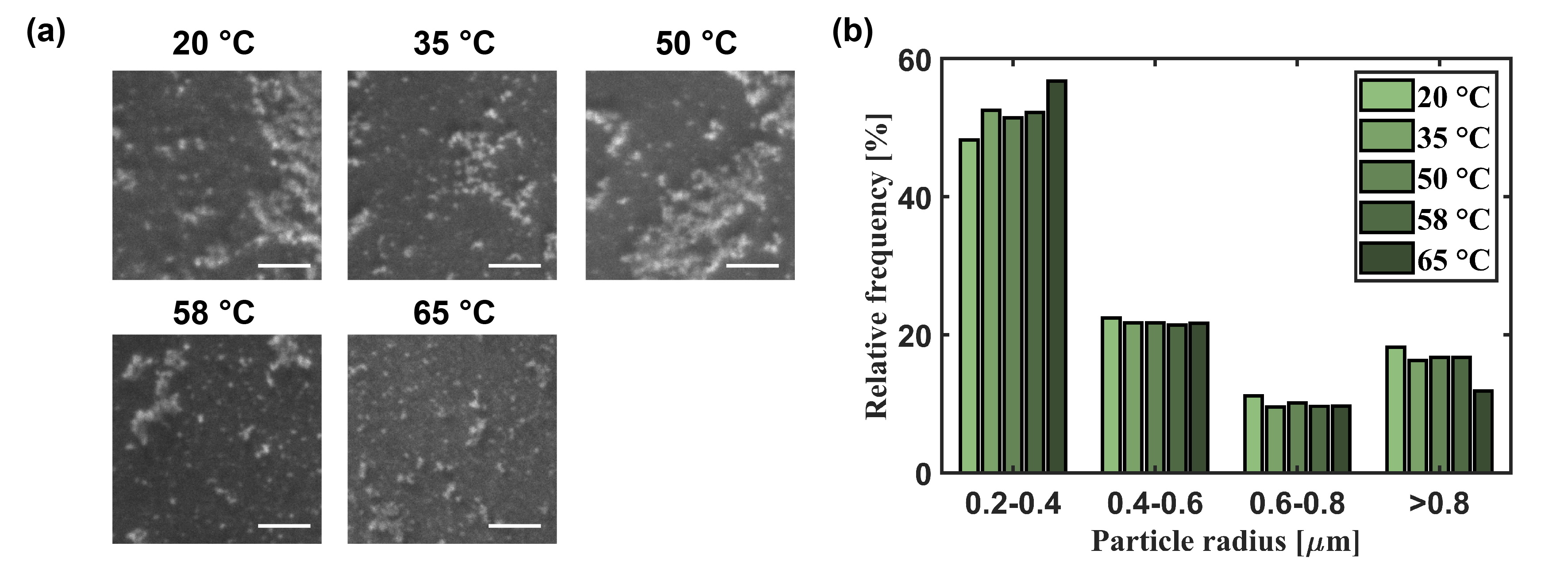}
\renewcommand{\captionfont}{\linespread{1.6}\normalsize}
\caption{(a) TIRF images of asphaltene particles at the final state of different temperatures. The length of the scale bar is 5 $\mu m$. (b) Relative frequency of size distribution of asphaltene particles at different temperatures.}
\label{temp tirf}
\end{figure}

Figure \ref{temp tirf}(a) shows the TIRF images of asphaltene particles. The quantitative size distribution at the submicron scale is shown in Figure \ref{temp tirf}(b). In summary, PSMPs are less likely to aggregate into fractal particles with increasing temperature. 

Equations (\ref{beta_5}) and (\ref{alpha}) show both the collision frequency and efficiency increase with temperature. Therefore, the overall right-shift trend from 20 $^\circ C$  to 58 $^\circ C$  in micron scale is expected in Figure \ref{temp results}(c) and (d). The leftwards shift from 58 $^\circ C$ to 65 $^\circ C$ is caused by asphaltene dissolution in n-heptane at 65 $^\circ C$, as we discussed in the previous section. However, it is surprising that $R_p$ increases apparently from $\sim$ 48\% to $\sim$ 58\% with the increase of temperature at the submicron scale (Figure \ref{temp tirf}(b)). In this regard, it is to be noted that asphaltene is a mixture of several fractions. The responses of various fractions to temperature are different \cite{yang2016effect}. The increase in temperature causes some fractions of asphaltene become more soluble in n-heptane. The Hildebrand solubility parameter difference in Equation (\ref{beta_5}) is lower. The repulsive interaction between this 'soluble-like' fraction of asphaltene particles is high when they form PSMPs, causing a decrease in collision efficiency for further aggregation. In addition, a decrease in $SC$ also leads to less possibility of PSMPs aggregation \cite{meng2022size}. As a result, the PSMPs formed by these `soluble-like' asphaltene do not form aggregates but exist as PSMPs on the substrate to cause a higher $R_p$.




\section{Conclusions}
 Dilution-induced asphaltene precipitation was experimentally investigated using a solvent exchange method in a microchamber. The size distribution of asphaltene precipitates in high resolution images was analyzed and compared after 20 different mixing conditions with a paraffinic solvent, including flow rate, chamber height and orientation, and the solvent temperature. 
 In the same microfluidic chamber, increasing the flow rate of paraffinic solvents accelerated the aggregation rate but caused the particle growth to stop at smaller levels. However, the flow rate had no obvious effect on the surface coverage of asphaltene precipitates. The surface coverage and size of asphaltene precipitates increased with the chamber height, regardless of the orientation of the chamber. But both the surface coverage and size of asphaltene particles of the horizontally placed device were larger than the vertically placed device when the channel height was higher than 180 $\mu m$. The difference between a horizontally and a vertically chamber was attributed the formation of convective rolls caused by gravity, which vanished in a narrow chamber (90 $\mu m$ in height). Moreover, our results showed that  higher temperatures resulted in lower $SC$ and larger particle sizes from 20 to 65 $^\circ$C, possibly due to more stable asphaltene in the mixture liquid at higher temperature.  For all mixing conditions, the oversaturation pulse resulted from the solvent exchange process were simulated using a finite element method based multiphysics solver.  Our model results corroborated well with the experimental results. The combination of the experimental and simulation results demonstrated the significance of mixing dynamics on the morphology of the asphaltene precipitates.

The important implication from this study is that mixing conditions, such as flow rate, temperature, orientation, and dimension of microchamber, may be potentially leveraged as a cost-saving and effective approach to controlled asphaltene precipitation without using additional paraffinic solvents. \textcolor{black}{Our work covers a wide range of mixing conditions to be relevant to many situations.} Furthermore, this implication is not only applicable to asphaltene precipitation, but to other dilution-induced phase separation processes, such as nanoparticle aggregation or polymer precipitation.

\section*{Declaration of Competing Interests}
The authors declare that they have no known competing financial interests or personal relationships that could have appeared to influence the work reported in this paper.

\section*{Acknowledgement}
The authors acknowledge the funding support from the Institute for Oil Sands Innovation (IOSI) (project number IOSI 2018–03), from the Natural Science and Engineering Research Council of Canada (NSERC)-Collaborative Research and Development Grants, and from the Canada Research Chair Program and from Canada Foundation for Innovation, John R. Evans Leaders Fund. The authors are grateful for technical support from IOSI lab, particularly from Lisa Brandt and Brittany MacKinnon. We also thank Compute Canada (www.computecanada.ca) for the computational facility and technical support. We are also grateful to Gilmar F. Arends for his help on symmetry software, Dr. Shantanu Maheshwari for his help on COMSOL simulations, and our industry steward Dr. Sepideh Mortazavi Manesh for the fruitful discussion.

\section*{Nomenclature}

$\alpha_{i,j}$ = Collision frequency ($m^3/mol \cdot s$)

$\beta_{i,j}$ = Collision efficiency

$\delta$ = Hildebrand solubility parameter ($MPa^{1/2}$)

$C$ = Concentration ($mol/L$)

$d$ = Diameter of particle ($\mu m$)

$D$ = Diffusion coefficient ($m^2/s$) 

$F$ = Interfacial tension ($N/m$)

$g$ = Gravitational constant ($m/s^2$)

$G$ = Shear force ($s^{-1}$)

$h$ = Channel height ($\mu m$)

$J$ = Mass flux ($kg / m^{2} \cdot s^{}$)

$k_B$ = Boltzmann constant ($J/K$)

$K_{i,j}$ = Aggregation kernel

$K_m$ = Correction for discontinuity of fluid

$\mu$ = Viscosity ($mPa \cdot s$)

$\nu$ = Kinematic viscosity ($m^2/s$)

$n$ = quantity of particle

$N$ = Avogadro constant ($mol^{-1}$)

\textcolor{black}{$N_{psmp}$ = Quantity of primary submicron particles}

\textcolor{black}{$N_{smp}$ = Quantity of submicron particles}

$\phi$ = Level-Set function

$P$ = Pressure ($Pa$)

$Pe$ = P\'{e}clet number

$Q$ = Flow rate ($m^3/s$)

$\rho$ = Density ($kg/m^3$)

$R$ = Ideal gas constant ($J / K^{} \cdot mol^{}$)

$Ra$ = Rayleigh number

$Re$ = Reynolds number

$R_p$ = Ratio of primary submicron particles (\%)

$t$ = Time ($s$)

$T$ = Temperature ($K$)

$U$ = Mass average velocity ($kg/s$)

$w$ = Width ($\mu m$)

$\overline{X}$ = Particle moving distance ($m$)

\bibliography{ref}

\end{doublespace}
\newpage
\textbf{\Large{Graphical abstract}}
\begin{figure}[ht]
	\centering
	\includegraphics[width=\textwidth]{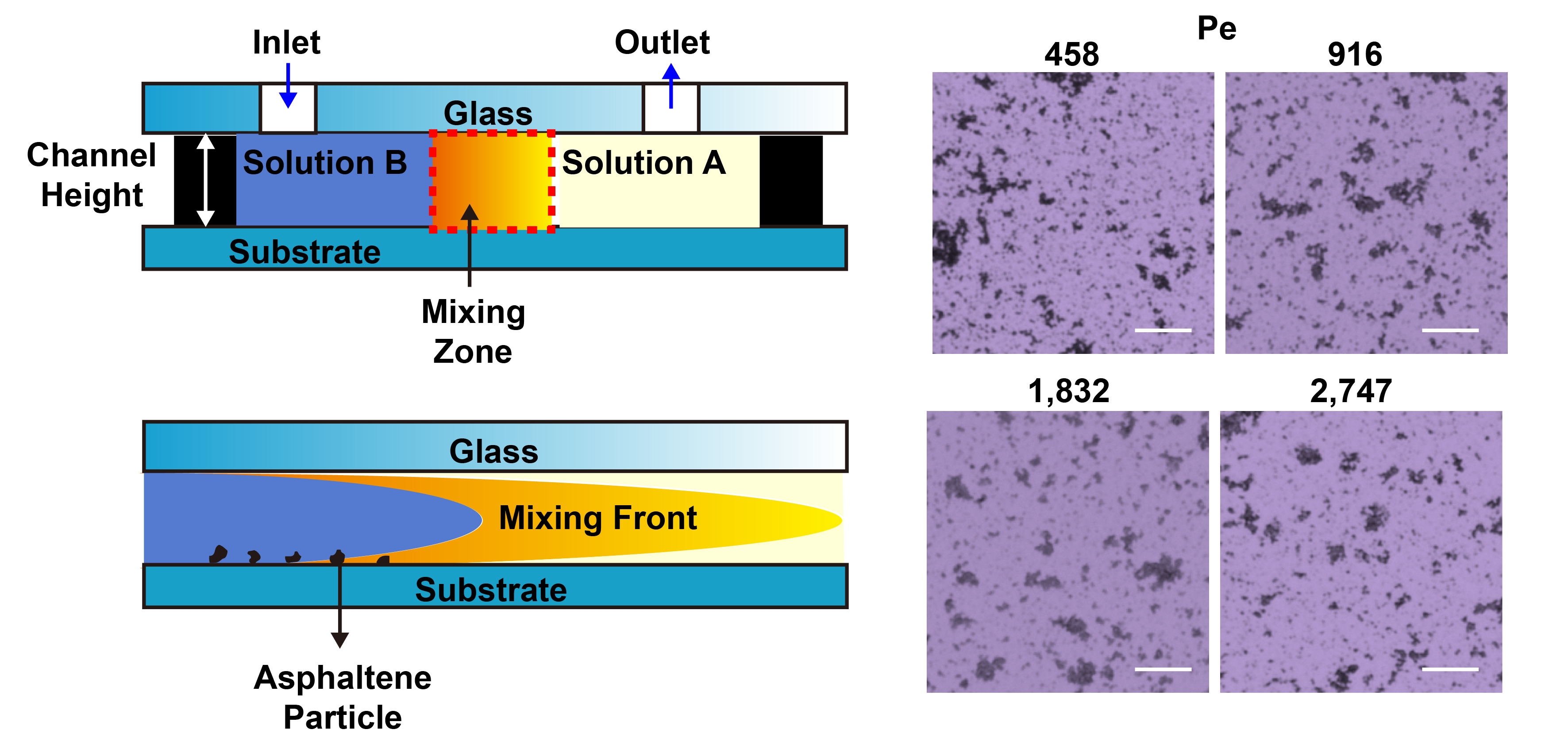}
\end{figure}

\textbf{\Large{Highlights}}
\begin{itemize}

\color{black}
\item Analyze asphaltene precipitates > 200 nm from mixing a model oil with heptane
\item Simulate mixing dynamics by a validated multiphysics model 
\item Reveal significance of mixing conditions in asphaltene precipitation 
\item Local concentration and shear forces impact size and amount of the precipitates
\item Primary submicron particles form universally under all 20 mixing conditions

\end{itemize}

\end{document}


\begin{doublespace}
\begin{frontmatter}

\title{\Large{Supporting information}\\}

\title{Asphaltene precipitation under controlled mixing conditions in a microchamber}

\author[First]{Jia Meng}
\author[First,Second]{Chiranjeevi Kanike}
\author[First]{Somasekhara Goud Sontti}
\author[Second]{Arnab Atta}
\author[First]{Xiaoli Tan\corref{cor1}}
\ead{xiaolit@ualberta.ca}
\author[First]{Xuehua Zhang\corref{cor2}}
\ead{xuehua.zhang@ualberta.ca}
\address[First]{Department of Chemical and Materials Engineering, University of Alberta, Alberta T6G 1H9, Canada}
\address[Second]{Department of Chemical Engineering, Indian Institute of Technology Kharagpur, Kharagpur, West Bengal 721302, India}
\cortext[cor1]{Corresponding author}
\cortext[cor2]{Corresponding author}

\end{frontmatter}


\renewcommand{\thefigure}{S\arabic{figure}}
\renewcommand{\thetable}{S\arabic{table}}





\section{Large asphaltene layer}
\begin{figure}[ht]
\centering
\includegraphics[width=1\columnwidth]{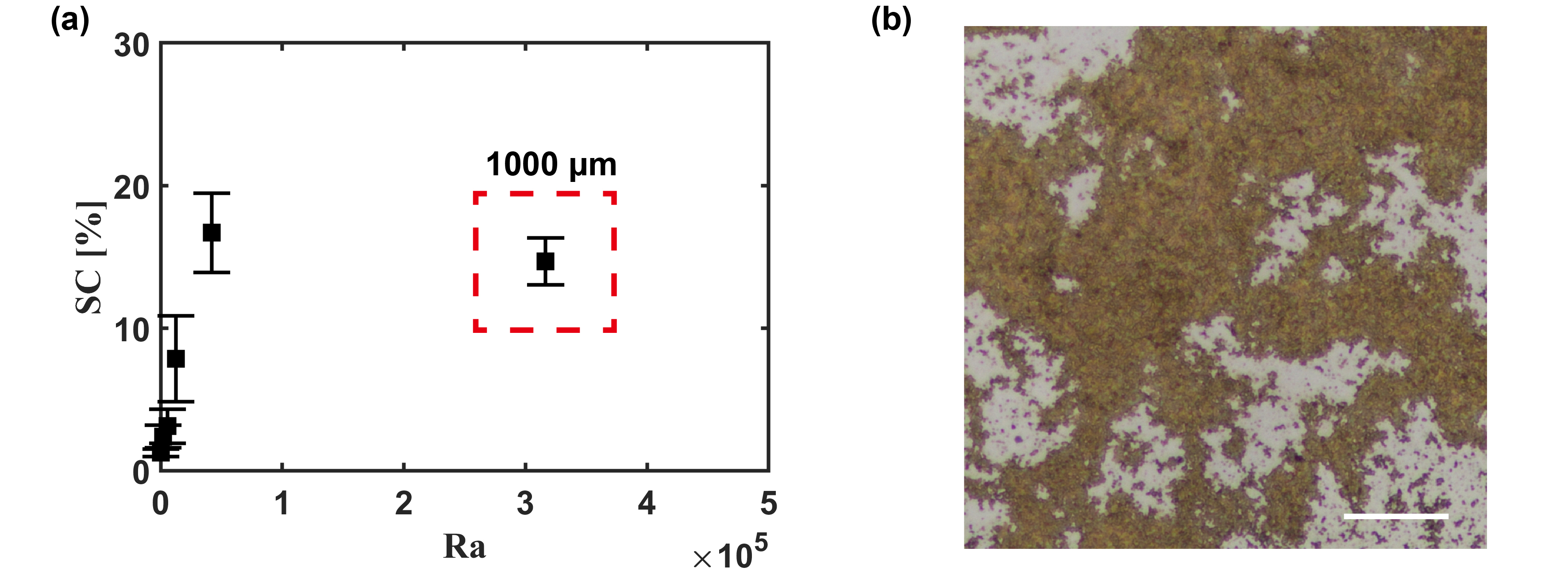}
\captionsetup{font={normal}}
\renewcommand{\captionfont}{\normalsize}
\renewcommand{\captionfont}{\linespread{1.6}\normalsize}
\caption{(a) SC of asphaltene at the final state as a function of Rayleigh number. The red-colored point is of 1,000 $\mu m$ channel height. (b) Optical image of layer of sediments in the 1,000 $\mu m$ channel height. The length of the scale bar is 50 $\mu m$.}
\label{1000}
\end{figure}

When the channel height of the microfluidic chamber is 1000 $\mu m$, we noticed the formation of large asphaltene layer, as shown in Figure \ref{1000}(b). The corresponding COMSOL simulation is shown in Figure \ref{COMSOL_1000}.

\begin{figure}[ht]
\centering
\includegraphics[width=\columnwidth]{COMSOL_1000.jpg}
\captionsetup{font={normal}}
\renewcommand{\captionfont}{\normalsize}
\renewcommand{\captionfont}{\linespread{1.6}\normalsize}
\caption{COMSOL simulations for the profile of the pulse of oversaturation in horizontally placed 1000 $\mu m$ microchannel.}
\label{COMSOL_1000}
\end{figure}

\begin{figure}[ht]
\centering
\includegraphics[width=1\columnwidth]{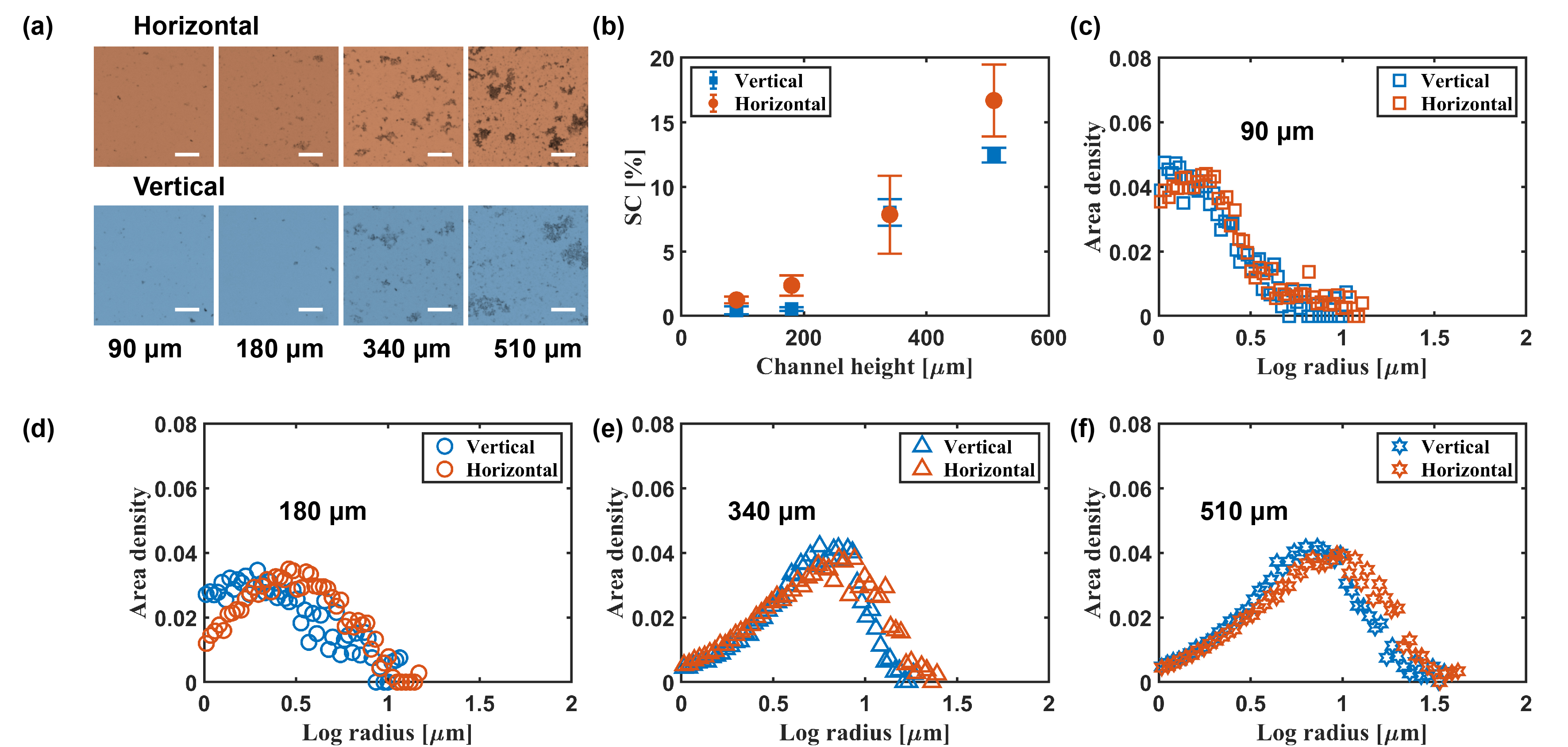}
\renewcommand{\captionfont}{\linespread{1.6}\normalsize}
\caption{(a) Optical images of asphaltene at the final state in vertical and horizontally placed devices. The channel heights are 90 $\mu m$, 180 $\mu m$, 340 $\mu m$, and 510 $\mu m$ from left to right. The length of the scale bar is 20 $\mu m$. The images are false-colored. (b) Comparison of $SC$ of vertically and horizontally place devices. Size distribution of asphaltene particles in vertically and horizontally placed devices with channel heights of (c) 90 $\mu m$, (d) 180 $\mu m$, (e) 340 $\mu m$, and (f) 510 $\mu m$.}
\label{vertical}
\end{figure}

\section{COMSOL simulation parameters at different temperatures}
Density and viscosity of n-heptane and toluene at different temperatures were obtained from the literature, as summaried in Table \ref{parameters}.

\begin{table}[ht]
\centering
\renewcommand{\captionfont}{\linespread{1.6}\normalsize}
\caption{Density and viscosity of n-heptane and toluene at different temperatures}
\label{parameters}
\begin{tabular}{|l|ll|ll|}
\hline
 & \multicolumn{2}{l|}{Density ($g/cm^3$)} & \multicolumn{2}{l|}{Viscosity ($mPa \cdot s$)} \\ \hline
 & \multicolumn{1}{l|}{n-Heptane} & Toluene & \multicolumn{1}{l|}{n-Heptane} & Toluene \\ \hline
20 $^\circ$C  & \multicolumn{1}{l|}{0.684 \cite{dandekar1998measurement}} & 0.867 \cite{Engineering} & \multicolumn{1}{l|}{0.417 \cite{dandekar1998measurement}} & 0.590 \cite{santos2006standard} \\ \hline
35 $^\circ$C  & \multicolumn{1}{l|}{0.671 \cite{pena1978isothermal}}  & 0.853 \cite{Engineering} & \multicolumn{1}{l|}{0.348 \cite{sagdeev2013experimental}} & 0.494 \cite{santos2006standard} \\ \hline
50 $^\circ$C  & \multicolumn{1}{l|}{0.658 \cite{iwahashi1990dynamical}} & 0.839 \cite{Engineering} & \multicolumn{1}{l|}{0.315 \cite{iwahashi1990dynamical}} & 0.422 \cite{santos2006standard} \\ \hline
58 $^\circ$C  & \multicolumn{1}{l|}{0.649 \cite{sagdeev2013experimental}} & 0.832 \cite{Engineering} & \multicolumn{1}{l|}{0.275 \cite{sagdeev2013experimental}} & 0.390 \cite{santos2006standard} \\ \hline
65 $^\circ$C  & \multicolumn{1}{l|}{0.645 \cite{sagdeev2013experimental}} & 0.825 \cite{Engineering} & \multicolumn{1}{l|}{0.264 \cite{sagdeev2013experimental}} & 0.365 \cite{santos2006standard} \\ \hline
\end{tabular}
\end{table}

Mutual diffusion coefficients of toluene and n-heptane binary mixture at different temperatures were calculated using the same method reported in our previous study \cite{meng2022}. In summary, the diffusion coefficient was estimated by Darken model \cite{almasi2021molecular,zhu2016prediction}:

\begin{equation}
    D_{ij} = \left(x_jD_{i}^{*} + x_iD_{j}^{*} \right) \Gamma
\label{equationDarken}
\end{equation}

The thermodynamics correction factor ($\Gamma$) was estimated based on activity coefficient ($\gamma_i$) as shown in Equation (\ref{equationthermo}). The activity coefficient was calculated by UNIFAC thermodynamic model with the software package Symmetry \cite{package_s}.

\begin{equation}
    \Gamma = 1 + x_i \left( \frac{\partial \ln \gamma_i}{\partial x_i} \right)
\label{equationthermo}
\end{equation}

The self-diffusion coefficients ($D_{1}^{*}$ and $D_{2}^{*}$) in the Darken model (Equation \ref{equationDarken}) were approximated using the Wilke-Chang equation:

\begin{equation}
    D_{i}^{*} = \frac{7.4 \times 10^{-8} \left(\psi_jM_j \right)^{0.5}T}{\mu_{j} \times V_{i,BP}^{0.6}}
\label{Wilke-Chang}
\end{equation}

$M_j$, $T$, $\mu_{j}$ and $V_{i,BP}$ represent the molecular weight of the solvent, temperature, solvent viscosity, and molar volume of solute at normal boiling point conditions, respectively. The estimated diffusion coefficients at different temperatures are summarized in Figure \ref{diffusion coefficient}.

\begin{figure}[ht]
\centering
\includegraphics[width=\columnwidth]{diffusion coefficient.jpg}
\captionsetup{font={normal}}
\renewcommand{\captionfont}{\normalsize}
\renewcommand{\captionfont}{\linespread{1.6}\normalsize}
\caption{Diffusion coefficients of toluene in n-heptane at (a) 20 $^\circ C$, (b) 35 $^\circ C$, (c) 50 $^\circ C$, (d) 58 $^\circ C$, (e) 65 $^\circ C$. Diffusion coefficient of toluene in n-heptane vs. temperature. }
\label{diffusion coefficient}
\end{figure}

\section*{Nomenclature}
$\gamma_i$ = Activity coefficient of solute at T and $x_i$

$x_i$ = Molar composition of solute $i$

$x_j$ = Molar composition of solvent $j$

$D_{i}^{*}$ = Self-diffusion coefficient of solute $i$ ($cm^2 \cdot s^{-1}$)

$D_{j}^{*}$ = Self-diffusion coefficient of solvent $j$ ($cm^2 \cdot s^{-1}$)

$D_{ij}$ = Mutual diffusion coefficient of solute $i$ in solvent $j$ ($cm^2 \cdot s^{-1}$)

$\Gamma$ = Thermodynamic factor

$M_j$ = Molecular weight of solvent $j$ ($g/mol$)

$T$ = Temperature ($K$)

$\mu_j$ = Viscosity of solvent $j$ at T ($cP$)

$\psi_j$ = Association factor of solvent $j$

$V_{i,BP}$ = Molar volume of solute $i$ at normal boiling temperature ($cm^3 \cdot mol^{-1}$)

\bibliography{sup}
\end{doublespace}